\documentclass[12pt,preprint]{aastex}
\usepackage{graphics,graphicx}
\usepackage{epstopdf}
\usepackage{epsfig}
\usepackage{amsmath}

\begin{document}
\def\etal {{\it et al. }}

\title{Nucleosynthetic Layers in the Shocked Ejecta of Cassiopeia A}
\author{Karl Isensee\altaffilmark{1}, Greg Olmschenk\altaffilmark{1}, Lawrence Rudnick\altaffilmark{1}, Tracey DeLaney\altaffilmark{2},
 Jeonghee Rho\altaffilmark{3}, J.D. Smith\altaffilmark{4},
William T. Reach\altaffilmark{5}, Takashi Kozasa\altaffilmark{6}, and Haley Gomez\altaffilmark{7}}

\altaffiltext{1}{Minnesota Institute for Astrophysics, University of Minnesota, Minneapolis,
MN 55455; isensee@astro.umn.edu, olmsc019@umn.edu, larry@astro.umn.edu}
\altaffiltext{2}{Department of Physics and Engineering, West Virginia 
Wesleyan College, 59 College Avenue, Buckhannon, WV 26201; 
delaney\_t@wvwc.edu}

\altaffiltext{3} {SETI Institute and SOFIA Science Center, NASA Ames Research Center, MS 211-1, Moffett Field, CA 94035; jrho@sofia.usra.edu}
\altaffiltext{4}{Ritter Astrophysical Observatory, University of Toledo,
Toledo, OH 43606 ; jd.smith@utoledo.edu}
\altaffiltext{5}{Stratospheric Observatory for Infrared Astronomy, Universities Space Research Association, MS 232-11, 
NASA Ames Research Center, Moffett Field, CA 94035;
wreach@sofia.usra.edu}
\altaffiltext{6}{Department of Cosmosciences, Graduate School of Science,
Hokkaido University, Sapporo 060-0810, Japan; kozasa@mail.sci.hokudai.ac.jp}
\altaffiltext{7}{School of Physics and Astronomy, Cardiff University, Queens Buildings, The Parade,
Cardiff, CF24 3AA, UK; haley.gomez@astro.cf.ac.uk}

\begin{abstract} We present a 3-dimensional analysis of the supernova remnant Cassiopeia~A using high resolution spectra from the Spitzer Space Telescope.  We observe supernova ejecta both immediately before and during the shock-ejecta interaction.  We determine that the reverse shock of the remnant is spherical to within 7\%, although the center of this sphere is offset from the geometric center of the remnant by 810 km s$^{-1}$.  We determine that the velocity width of the nucleosynthetic layers is $\sim$1000 km s$^{-1}$ over ~4000 square arcsecond regions, although the velocity width of a layer along any individual line of sight is $<$250~km~s$^{-1}$.  Si and O, which come from different nucleosynthetic layers in the progenitor star, are observed to be coincident in velocity space in some directions, but segregated by up to $\sim$500 km s$^{-1}$ in other directions.  We compare these observations of the nucleosynthetic layers to predictions from supernova explosion models in an attempt to constrain such models.  Finally, we observe small-scale, corrugated velocity structures that are likely caused during the supernova explosion itself, rather than hundreds of years later by dynamical instabilities at the remnant's reverse shock.

\end{abstract}

\keywords{ISM: Infrared -- ISM: Supernova Remnants -- ISM: Xray -- Supernovae: General -- Supernova Remnants: Cassiopeia A}

\section{Introduction} 

The supernova remnant Cassiopeia A (Cas~A) is a unique astrophysical laboratory due to its young age \citep[$\sim$340 years -][]{thor01, fesen06} and small distance \citep[only 3.4 kpc -][]{reed95}.  The remnant is just entering its Sedov-Taylor phase, so emission from both forward and reverse shocks can be detected \citep{hughes00}.  Emission at most wavelengths, including most of the infrared, is dominated by a $\sim$120$\arcsec$ radius ``Bright Ring''.  The Bright Ring is formed when supernova ejecta encounter Cas~A's reverse shock and are shocked, heated, and collisionally ionized.  It consists of undiluted ejecta rich in O, Si, S, Ne, Ar, Ca, and Fe \citep{chev78,douv99,hughes00,will03,hwl03,lhw03,morse04,ennis06}.

Studies of optical light echoes from the the explosion near peak light have led to the observation of weak hydrogen lines, indicating a supernova Type IIb origin for Cas~A \citep{krause08}.  Cas~A's progenitor was therefore a red supergiant that had lost most, but not all, of its hydrogen envelope.  X-ray studies indicate a total ejecta mass of $\sim$2M$_{\odot}$ \citep{will03}. If one adds to this the mass of the central compact object \citep{chak01}, Cas~A's progenitor had a total mass of at least 4M$_{\odot}$ immediately before the supernova explosion.  The estimated oxygen mass indicates a main sequence mass of $\sim$15-25M$_{\odot}$ \citep{young06, vink96}.

Although Cas~A's appearance is dominated by recently shocked ejecta, it also contains emission that is not the result of collisional ionization at the reverse shock, but photoionization by UV and X-ray emission from the shocked ejecta \citep{hs84, hf98, smith09}.  This material is seen toward the central region of the remnant at low radio frequencies \citep{kas95} and in the infrared \citep{rho08, smith09, delaney10, isens10}, and was demonstrated to be at lower densities and ionization state than recently shocked material on the Bright Ring through a combination of Doppler analysis and line ratio measurements \citep{smith09}.  These ejecta are often referred to as ``unshocked ejecta'' since they have yet to encounter the remnant's reverse shock.  That is not an accurate label, since Cas~A's forward shock and a reverse shock interacted with the ejecta during the supernova explosion itself. 


\subsection{Previous 3D Maps}

3D Maps of Cas~A have been made in the optical, infrared, and X-ray.  Doppler reconstructions in the optical used S and O emission lines \citep{law95,reed95} and showed that ejecta on the Bright Ring lie on a roughly spherical shell but are not uniformly distributed on that shell - most of the ejecta lie nearly in the plane of the sky.  They also observe that the center of the sphere is offset from the geometrical center of the spherical shell by $\sim$0.36pc along our line of sight.  This indicates that the ejecta are not traveling at the same velocity in all directions, which is consistent with previous results which indicated an asymmetric expansion for the ejecta \citep[e.g.][]{braun87, will02}.  These 3D reconstructions give us a selective snapshot of ejecta because only material that has recently encountered the remnant's reverse shock will emit strongly in the optical.

\cite{delaney10} created a 3D infrared and X-ray map of Cas~A from a \emph{Spitzer Space Telescope} spectral cube\footnote{Movies showing this 3D structure are available at http://chandra.harvard.edu/photo/2009/casa2/animations.html}.  \cite{isens10} used a similar IR data set, but at higher spectral resolution, to make a 3D map of ejecta in the center of the remnant.  The advantage of these IR maps lies in the fact that much of the ejecta in the IR will be detectable both before and after they interact with the reverse shock.  Both studies found a similar distribution of ejecta to that seen in the optical where the center of expansion is offset from the geometrical center of the remnant both in projection and along the line of sight.  These works were able to study the relationship of several nucleosynthetic layers and are discussed in the next section.

\subsection{Separation of Nucleosynthetic Layers}

Si and O emission are observed to be co-located in most regions \citep[e.g.][]{ennis06} in both the X-ray and infrared.  This indicates that the two layers have very similar velocities (less than 80 km~s$^{-1}$ difference).  However, evidence of layer differentiation is found in some directions in the X-ray \citep[e.g.][]{hughes00}, the optical \citep{fesen06}, and the IR \citep[e.g.][]{delaney10,isens10}, which was likely caused by the different layers of the star being ejected at different velocities in those directions, thus encountering the remnant's reverse shock at different times.

It should be emphasized that we can only observe mixing or separation in \emph{velocity space}.  We can easily detect any velocity gradients in the supernova explosion since we can detect Doppler velocities of $<$100~km~s$^{-1}$ in the IR, while typical observed velocities and velocities predicted by models are an order of magnitude larger \citep[e.g.][]{hammer10}.  However, we cannot detect any initial spatial separation of the nucleosynthetic layers - simulations predict that the relevant nucleosynthetic layers will be $<$ $10^{11}$ cm thick prior to the explosion \citep[e.g.][]{jog09}, but the typical ejecta clump size of $<$1$\arcsec$ corresponds to $\sim$ $10^{16}$ cm at Cas~A's distance.  Therefore, we cannot differentiate between a situation where two nucleosynthetic layers were separated during the supernova explosion but ejected at the same velocity, and one where the two layers were completely mixed during the explosion and ejected at the same velocity.  But, if we observe two layers that are currently separated in velocity space, we know that they were separated during the supernova explosion itself because, to the best of our knowledge, there is no mechanism that will impart different velocities to spatially overlapping elements.


\subsection{Geometrical Asymmetries}

Supernova explosion models predict substantial asymmetries due to effects such as rotation as well as instabilities \citep[e.g.][]{blondin03, burrows07, hammer10}.  Observations of both supernovae and supernova remnants have confirmed this picture.  Spectropolarimetric observations of unresolved supernovae have shown that all observed core collapse supernovae contain intrinsic polarization, indicating that there is a departure from spherical symmetry \citep{whe05}.  Although an axis-symmetric geometry, probably induced by jets, can be used to explain some features in some core collapse supernovae, significant departures from axial symmetry are needed to explain most observations \citep{wang08}.

IR 3D maps of supernova ejecta in Cas~A have found major asymmetries, both on global scales \citep{delaney10} and for smaller subsets of ejecta in the the entire supernova remnant \citep{isens10}.   These asymmetries are not immediately apparent in the visual appearance alone because the highly spherical reverse shock creates a large selection effect in that we can only observe ejecta near the shock at most wavelengths.

In this paper, we present an analysis of high spectral resolution Spitzer mappings of the ejecta on the Bright Ring of Cas~A.  This data set is an extension of that used by \cite{isens10}, and it contains regions with both recently shocked and interior ejecta.  In $\S$2 we present the observations and discuss the methods used in our analysis.  We describe those results in $\S$3 and discuss the physical implications in $\S$4.

\section{Observations and Analysis}

\begin{figure*}
    \centering
    \includegraphics[width=0.4\textwidth]{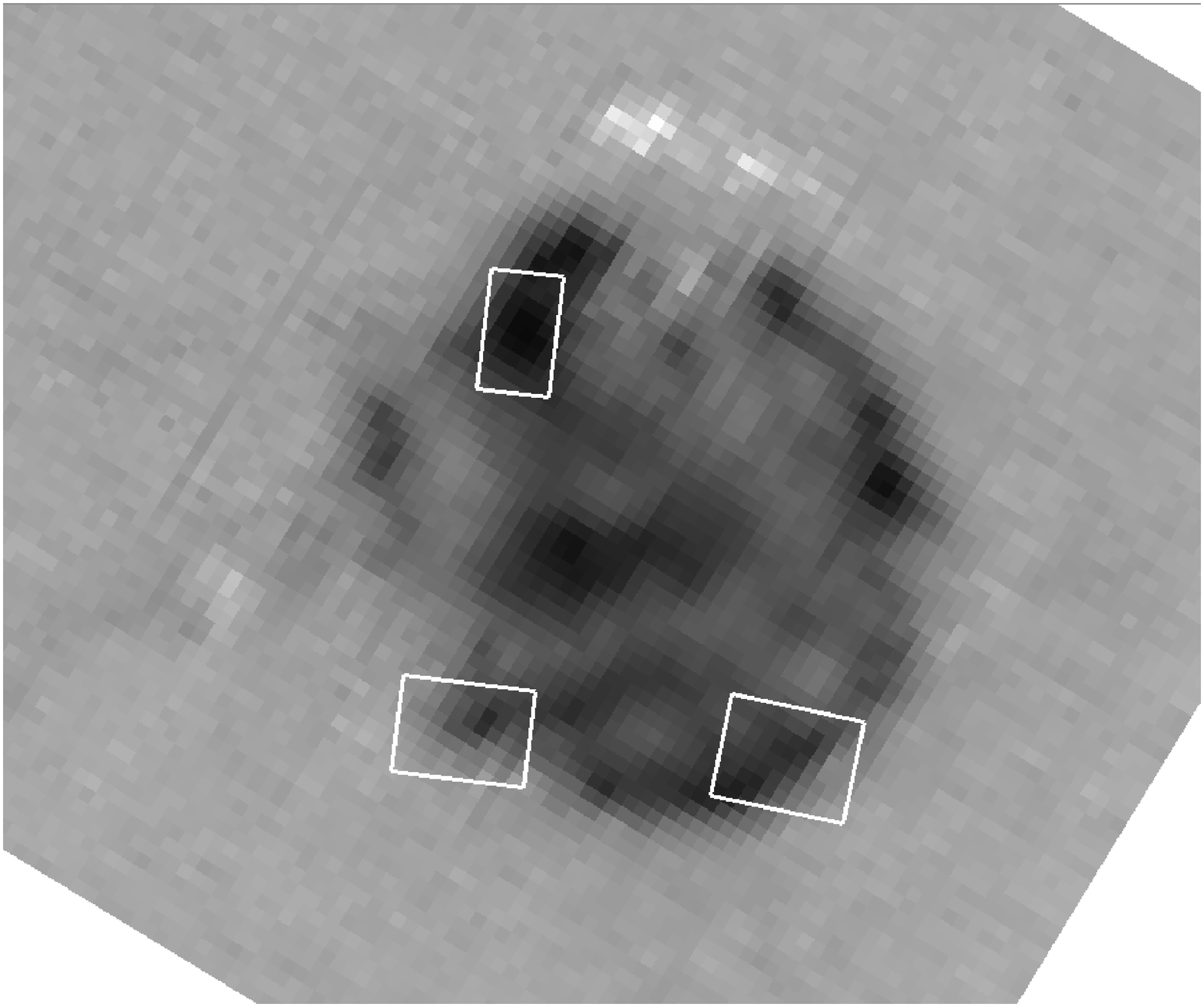}
    \includegraphics[width=0.4\textwidth]{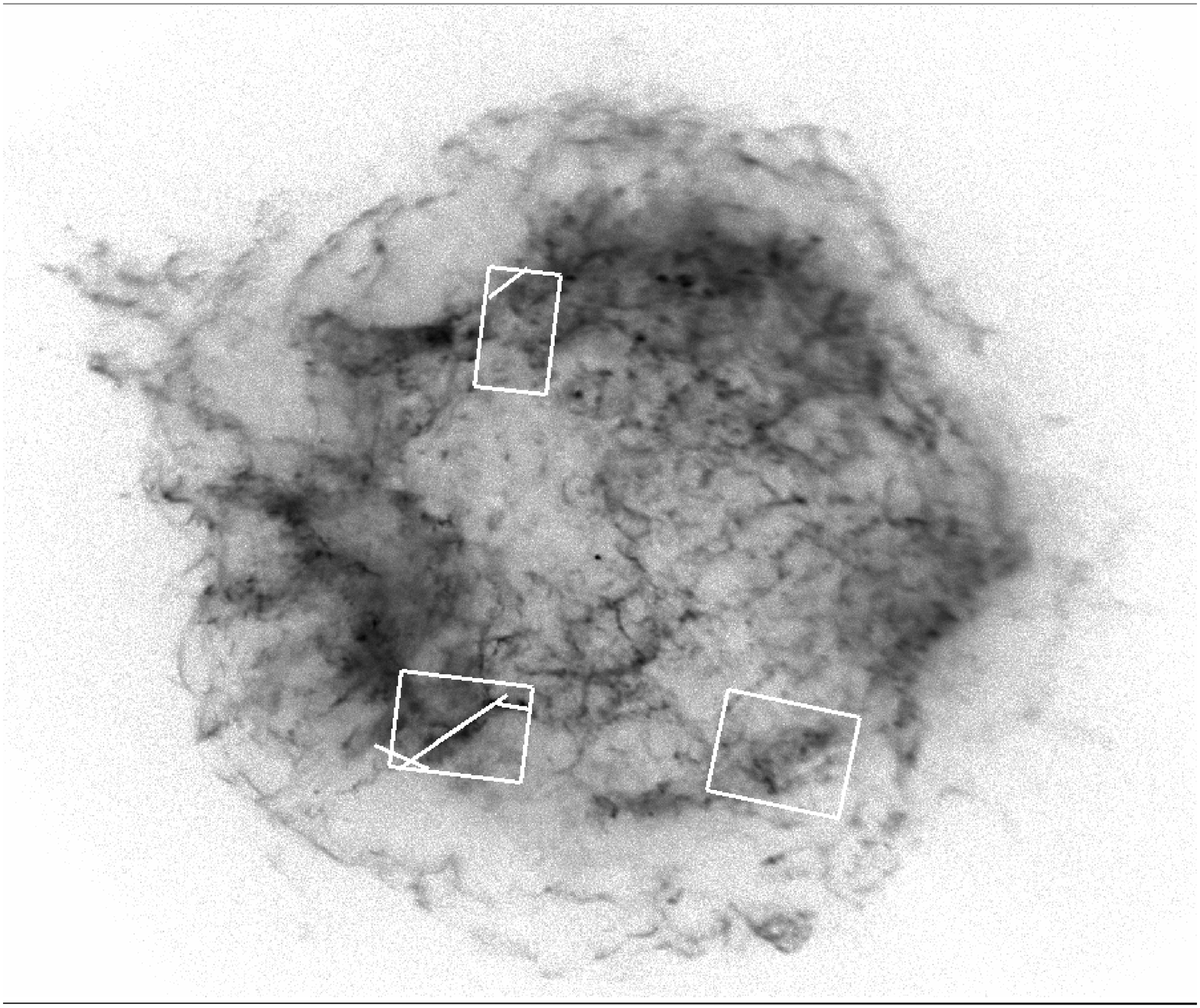}

\caption{\small 34.81$\mu$m [Si~II] \emph{Spitzer} IRS map (left) and X-ray Si \emph{Chandra} map (right) of Cas~A.  Both maps have been continuum subtracted.  The regions of high resolution data discussed in this text are indicated by the boxes.  The planes shown in Figure \ref{fig:xrayplanes} are indicated by the straight lines in the X-ray image.
\normalsize }
    \label{fig:xrayandir}
\end{figure*}

The \textit{Spitzer} Infrared Spectrograph (IRS) was used on August 30, 2007 to create spectral maps of select relatively bright regions in Cas~A whose locations are shown and labeled in Figure \ref{fig:xrayandir}.  High-resolution spectra (R$\sim$600 for all wavelengths) were taken from 20-35$\mu$m using the Long High (LH) module in all regions and from 10-35$\mu$m using both the Long High and Short High (SH) modules in the Southwest region.  The FWHM of unresolved spectral features in these observations is about 0.06$\mu$m at 35$\mu$m and about 0.02$\mu$m at 13$\mu$m.  The LH data were taken using a 61 second exposure at each position while the SH data were taken using a 31 second exposure at each position.  The pixel scale of the observations is $\sim$1.25$\arcsec$ and $\sim$2.5$\arcsec$ for the SH and LH modules respectively.  The background, which was taken from 3 separate 61 second observations adjacent to the remnant, was subtracted and 3D cubes were created using the S19 version of the IRS pipeline and the CUBISM software package \citep{smith07}.  The uncertainties in flux for each line of sight were calculated from the IRS pipeline using standard error propagation of the BCD level errors.

The undersampling of the IRS modules limits our uncertainties.  This is a systematic error that exists in the wavelength calibration data themselves, and is worst at the short-wavelength end of both the SH and LH modules.  Our obtainable wavelength accuracy is limited to roughly 1/2 of a spectral bin, or about 100 km~s$^{-1}$.  The relative wavelengths for a given line can be measured with higher accuracy from position to position or within multiple Doppler components in a given position.


\subsection{Spectra}

\begin{figure*}
    \centering
    \includegraphics[width=1.0\textwidth]{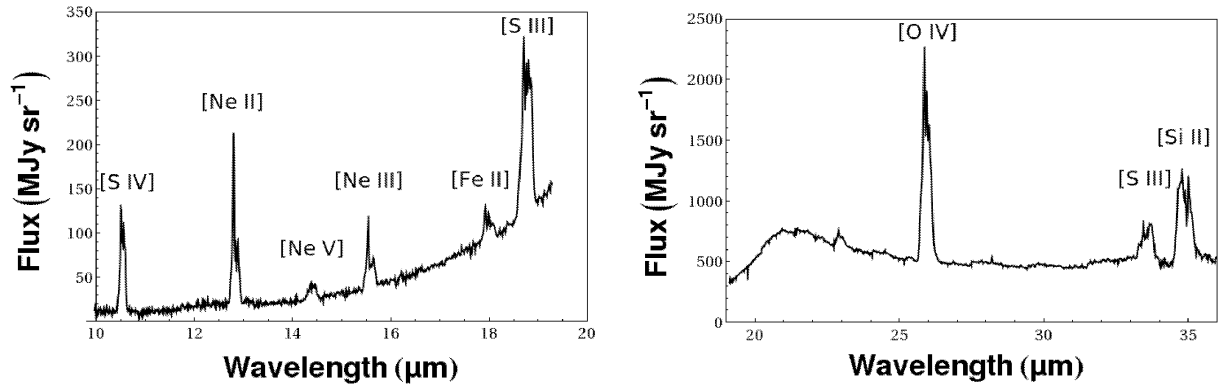}

\caption{\small Typical spectra from the SH and LH \emph{Spitzer} IRS module of emission in Cas~A.  The small bump near 23$\mu$m may be from the 22.9$\mu$m [Fe~III] line.
\normalsize }
    \label{fig:irspec}
\end{figure*}

Cas~A's infrared spectrum is dominated by bright emission lines as shown in Figure \ref{fig:irspec}.  The LH observations contain lines from [O~IV] at 25.9$\mu$m, [S~III] at 33.48$\mu$m, and [Si~II] at 34.81$\mu$m.  We tentatively identify the line near 23$\mu$m as the 22.9$\mu$m [Fe~III] line.   The lines observed in the LH module typically have peak fluxes from 100-10,000 MJy~sr$^{-1}$, with an rms noise of $\sim$20 MJy~sr$^{-1}$.  The SH observation contains lines from Ne, S, and Fe.  Typical peak fluxes are $\sim$300 MJy~sr$^{-1}$.


\subsection{Doppler Deconvolution}

We performed a Doppler deconvolution of the spectral lines for each line of sight from each ion separately using a spectral CLEAN algorithm \citep{ding99}.  We determined the uncertainties in Doppler velocity for each Doppler component by applying the spectral CLEAN to synthetic line data with a realistic range of signal to noise ratios, and using line free data in order to model the noise.  For both procedures, we use identical techniques to those described in \cite{isens10}.

Using synthetic data, the uncertainty in velocity for a single, isolated Doppler component was determined to be $<$25 km~s$^{-1}$, however, we could not differentiate two components from one another along the same line of sight that were within 65 km~s$^{-1}$ of one another.  Therefore, uncertainties in the absolute velocities are limited by the systematic errors in the calibration of $\sim$100 km~s$^{-1}$ rather than random uncertainties. 

\begin{figure*}
    \centering
    \includegraphics[width=0.85\textwidth]{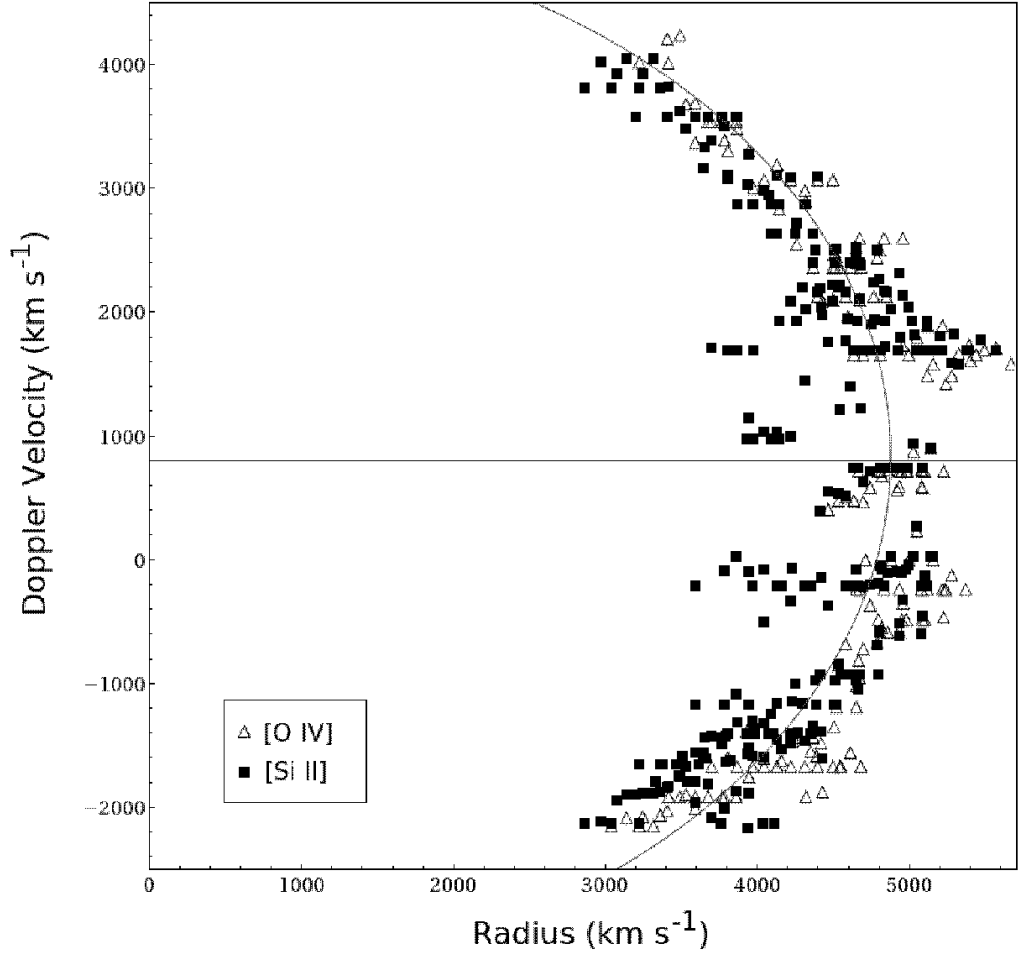}

\caption{\small Velocity vs radius plot for Si (filled squares) and O (open triangle).  Our assumption of no deceleration is good since all the ejecta lie nearly on the same semi-circle.  Note that the semi-circle is offset from 0 velocity by 810 km s$^{-1}$ as indicated by the solid horizontal line.
\normalsize }
    \label{fig:vvsr}
\end{figure*}

As in previous studies \citep[e.g.][]{reed95, delaney10, isens10}, we assume that the ejecta have been freely expanding at a constant velocity in order to determine their spatial coordinate perpendicular to the plane of the sky.  \cite{delaney10} demonstrate that this is a good assumption for IR emission by showing that nearly all ejecta plotted on a Velocity vs. Radius plot fall on a semi-circle.  We make a similar Velocity vs. Radius plot in Figure \ref{fig:vvsr} from our data set and find that ejecta in these regions fall on a semi-circle that is consistent with that found by \cite{delaney10}.  The assumption of constant velocity is still valid despite the fact that the ejecta were likely decelerated during the supernova explosion itself.  The subsequent behavior after the explosion is essentially identical to free expansion at a reduced velocity because any deceleration happened near t=0, z=0 where z is the spatial coordinate perpendicular to the plane of the sky.

\section{Results}

\subsection{3D Maps}

\begin{figure*}
    \centering
    \includegraphics[width=0.3\textwidth]{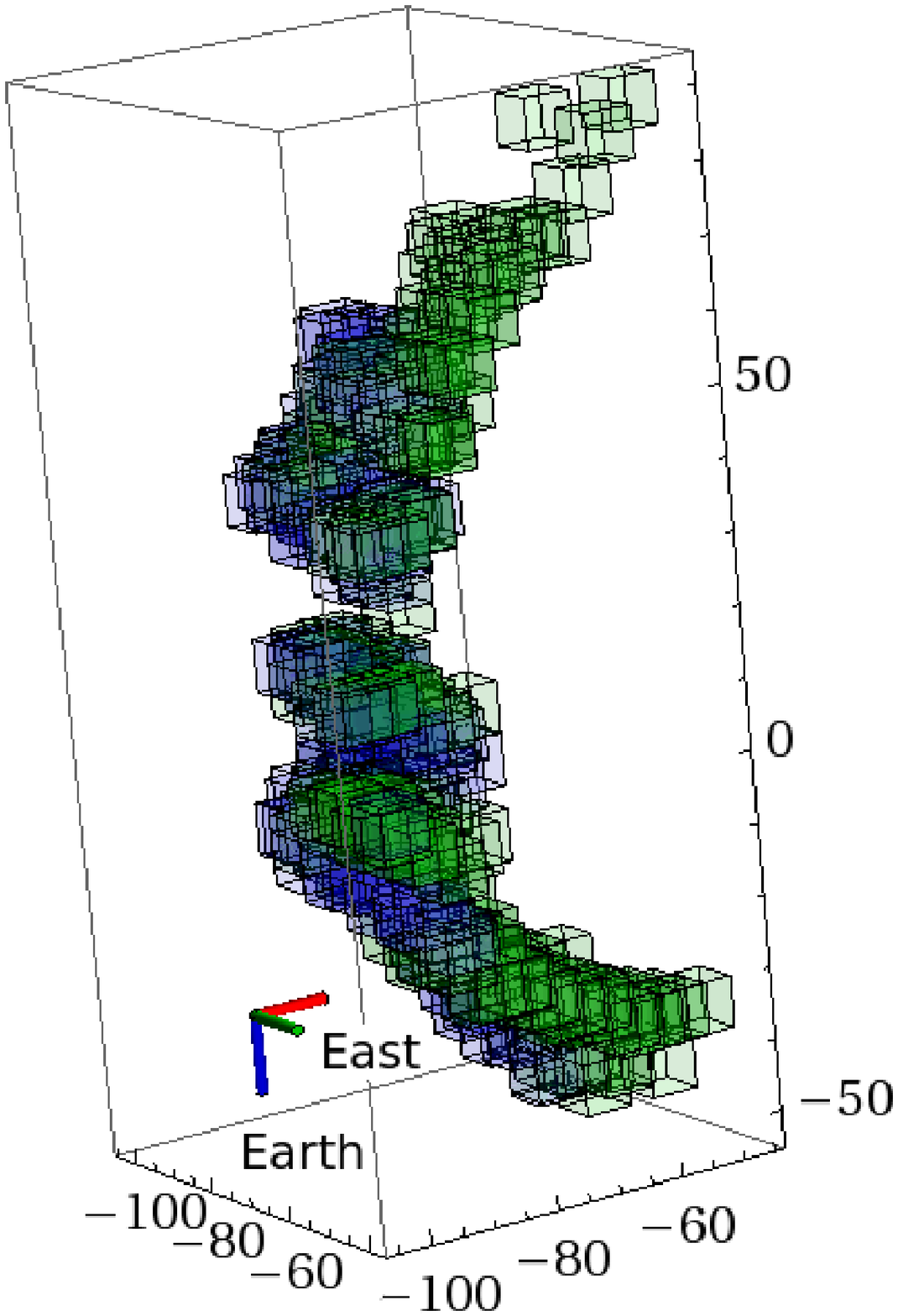}
    \includegraphics[width=0.3\textwidth]{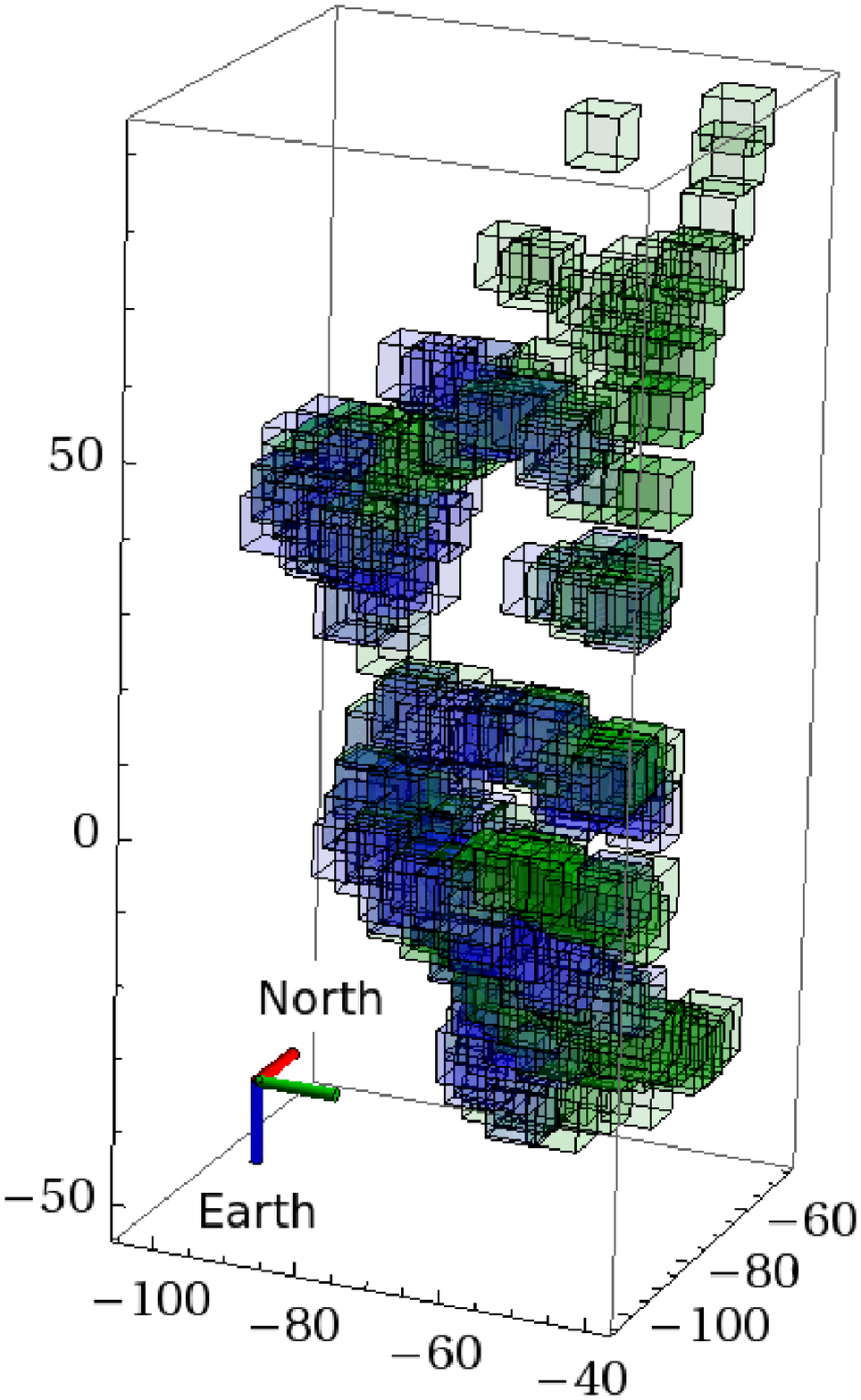}
    \includegraphics[width=0.3\textwidth]{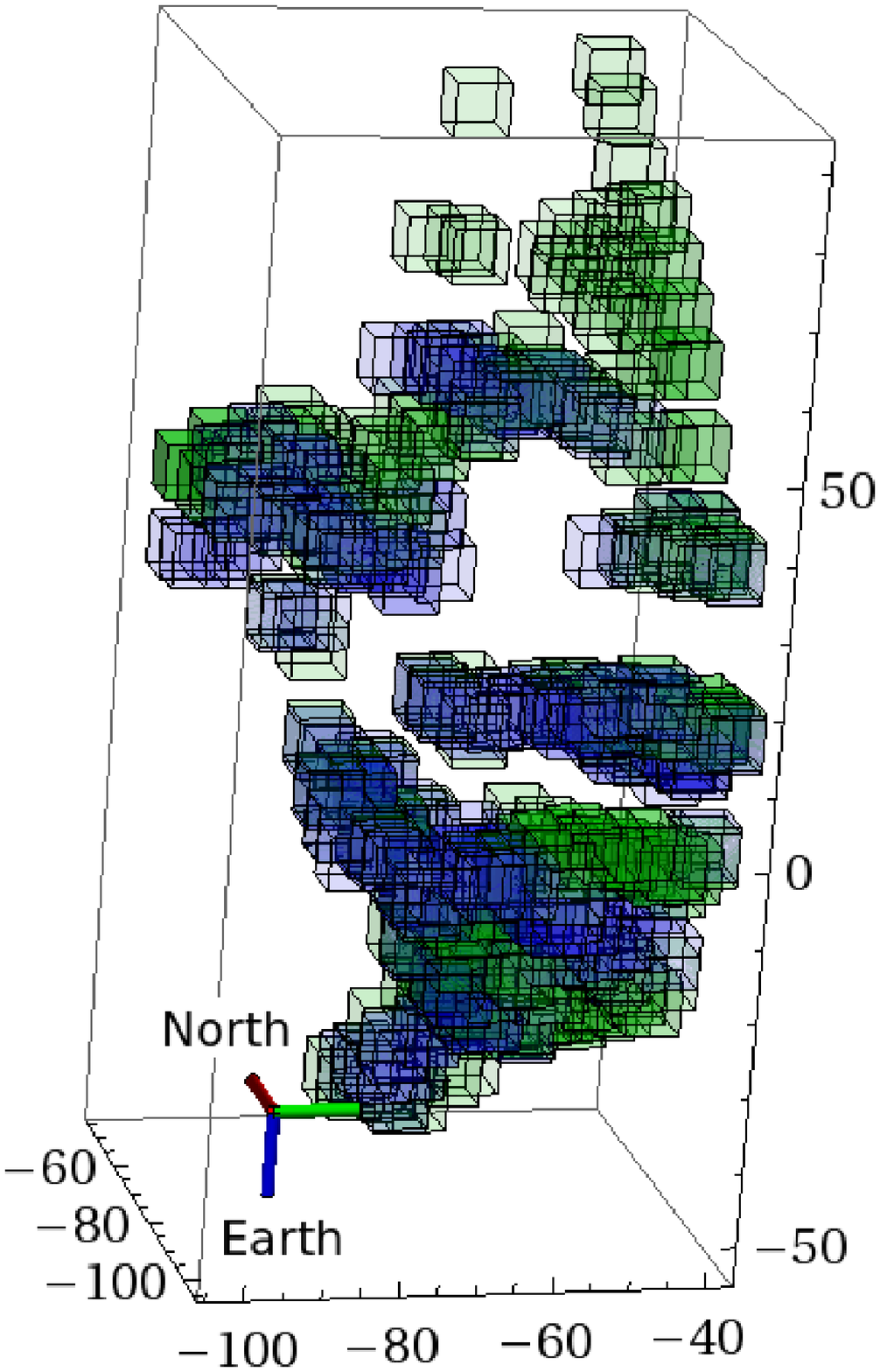}

\caption{\small 3D plot of the the 25.9$\mu$m [O~IV] line (blue) and 34.8$\mu$m [Si~II] line (green) in the Southwest of the remnant as viewed from three different angles.  The angles were chosen to best highlight the 3D structure of the ejecta.  The units on the axes are arcseconds from the center of expansion of the remnant.
\normalsize }
    \label{fig:sw}
\end{figure*}

\begin{figure*}
    \centering
    \includegraphics[width=0.45\textwidth]{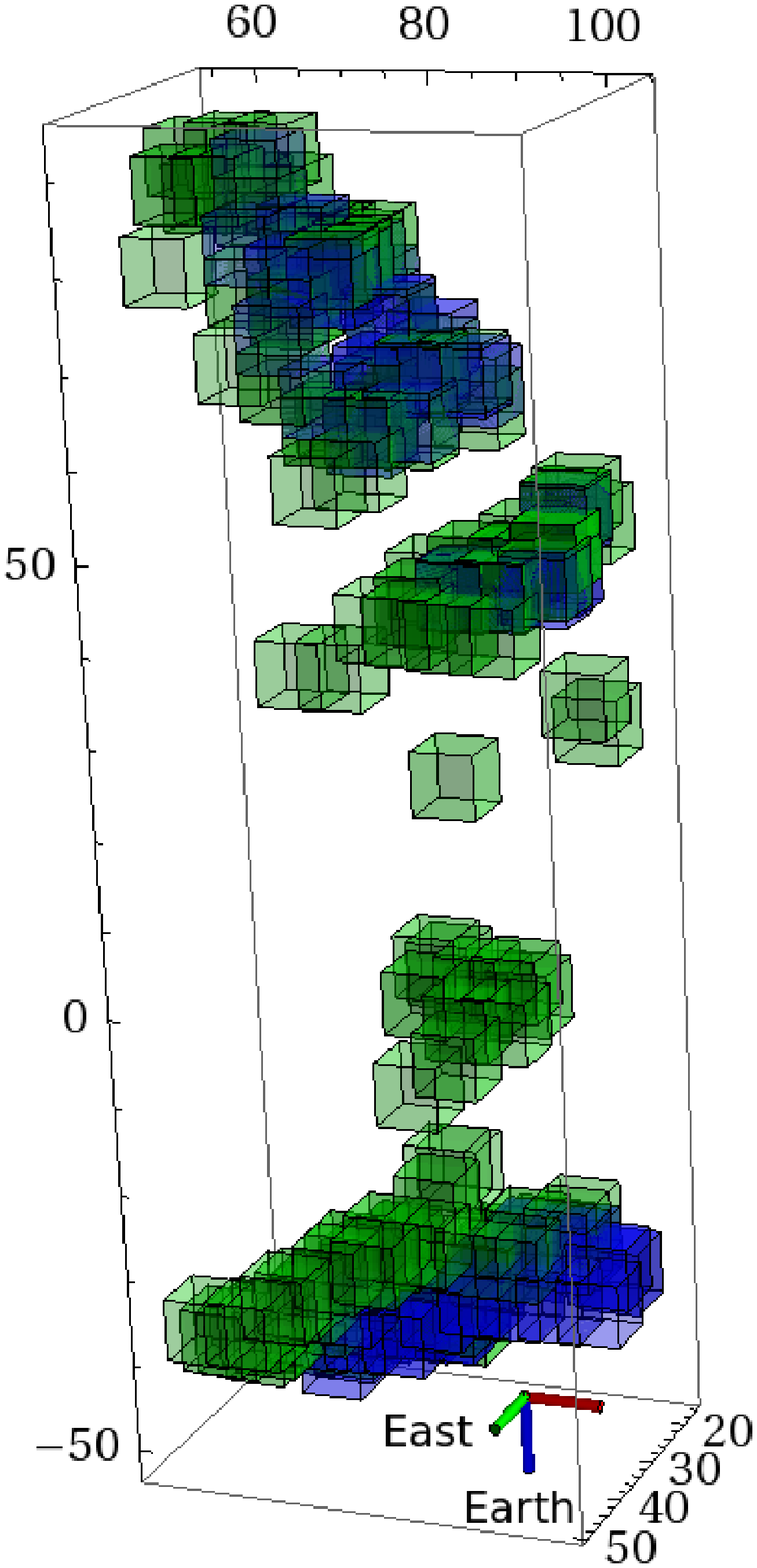}
    \includegraphics[width=0.45\textwidth]{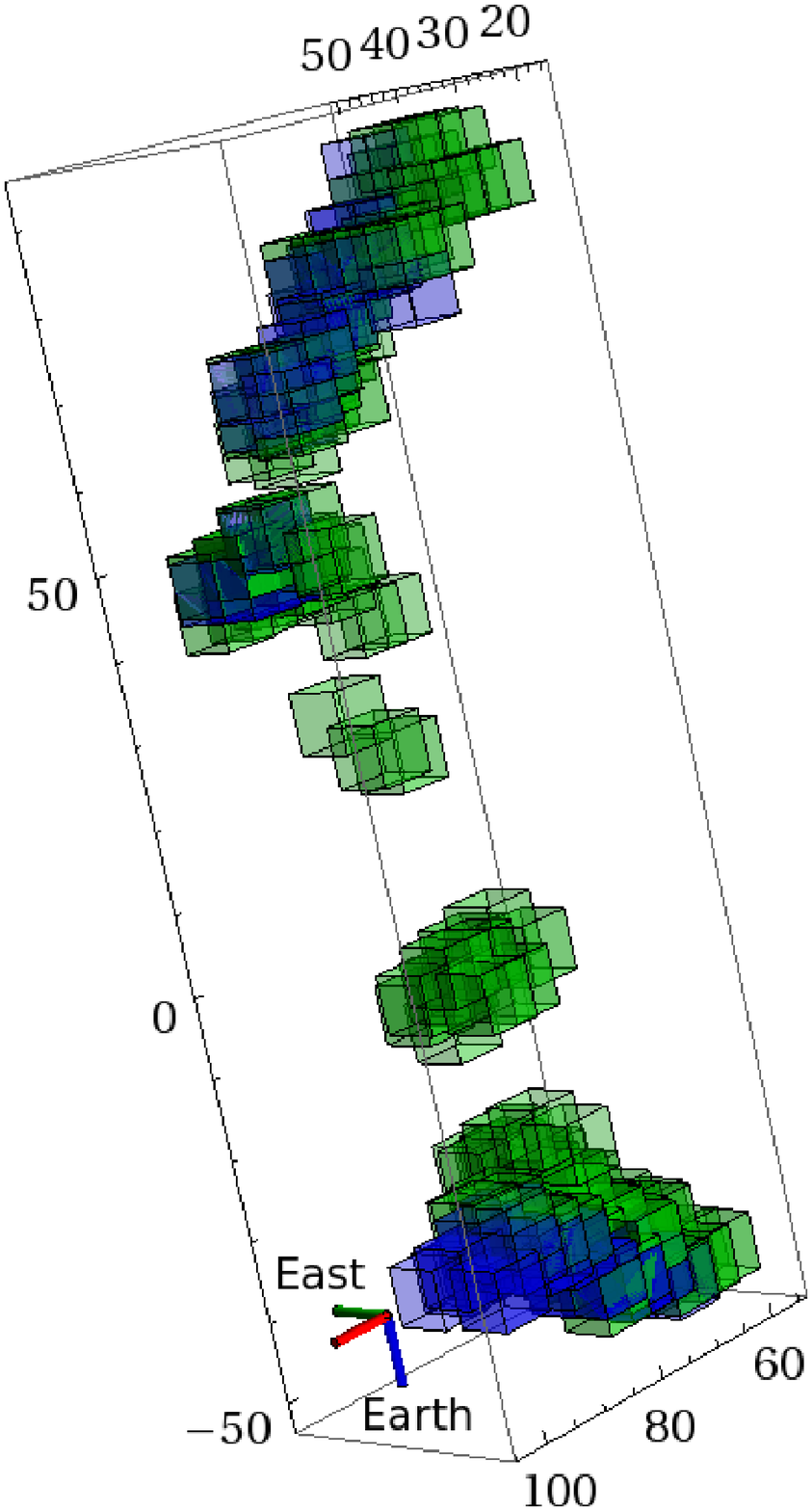}

\caption{\small 3D plot of the the 25.9$\mu$m [O~IV] line (blue) and 34.8$\mu$m [Si~II] line (green) in the Northeast of the remnant as viewed from two different angles.  The angles were chosen to best highlight the 3D structure of the ejecta.  The units on the axes are arcseconds from the center of expansion of the remnant.
\normalsize }
    \label{fig:ne}
\end{figure*}

\begin{figure*}
    \centering
    \includegraphics[width=0.45\textwidth]{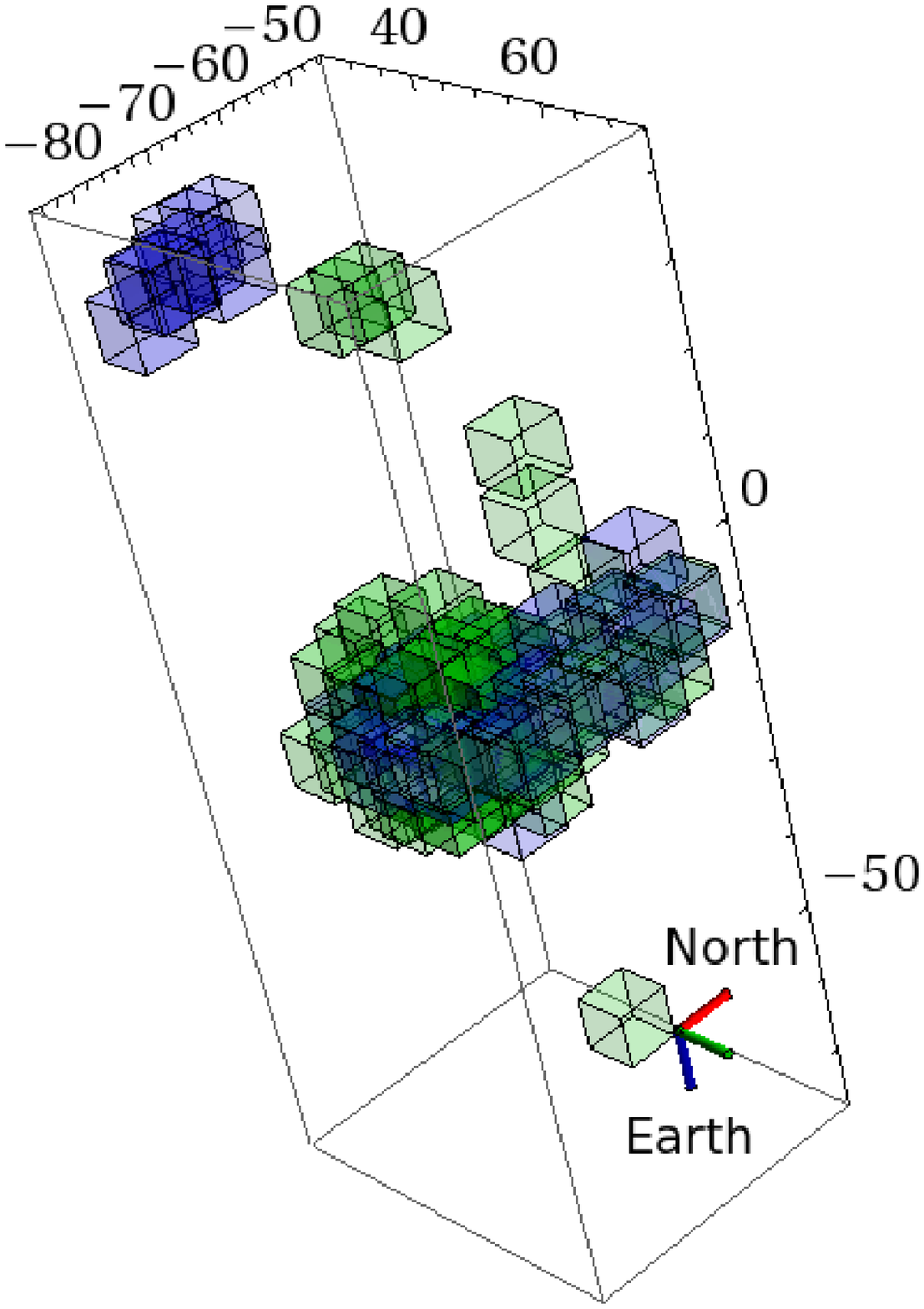}
    \includegraphics[width=0.45\textwidth]{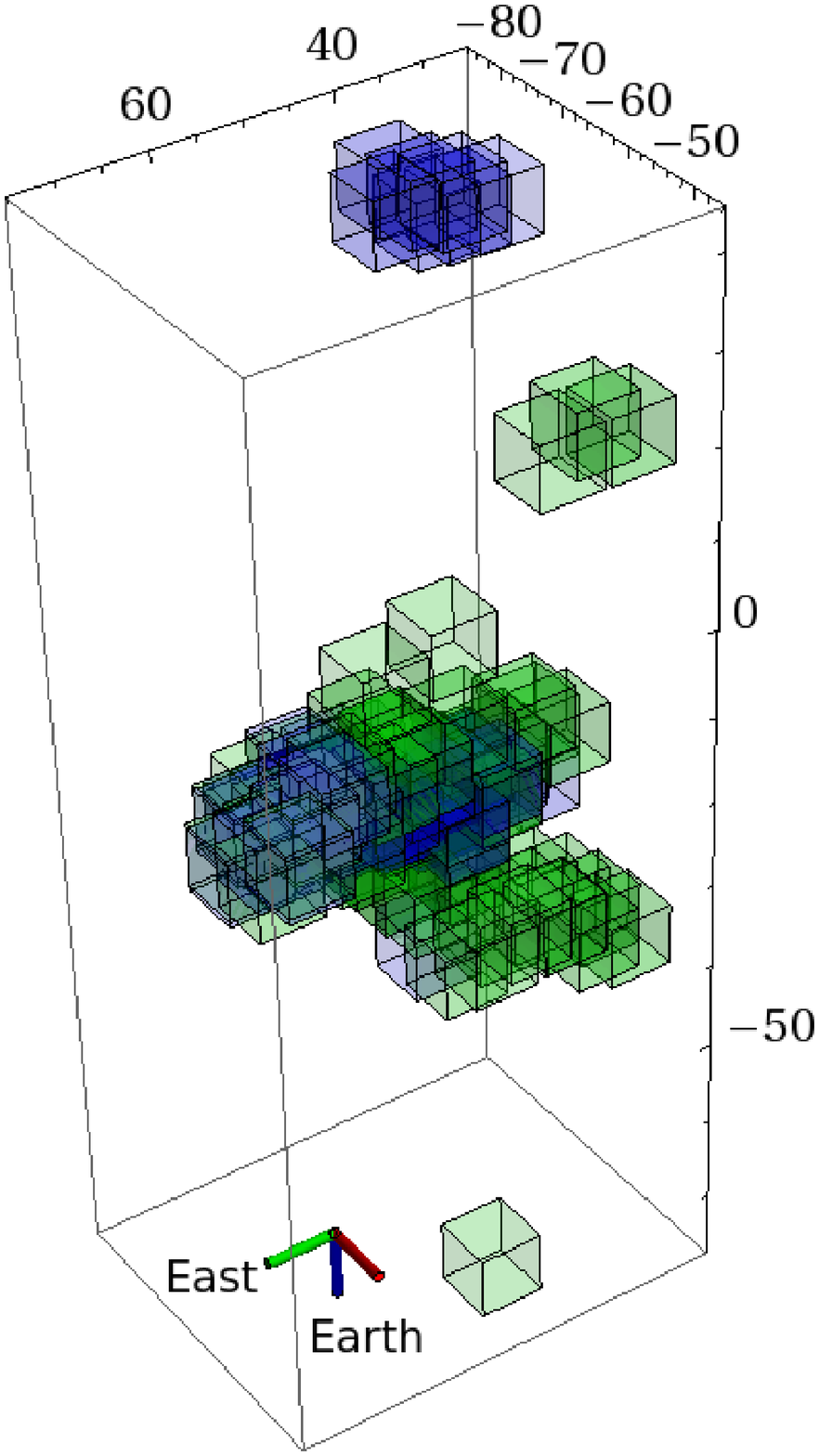}

\caption{\small 3D plot of the the 25.9$\mu$m [O~IV] line (blue) and 34.8$\mu$m [Si~II] line (green) in the Southeast of the remnant as viewed from two different angles.  The angles were chosen to best highlight the 3D structure of the ejecta.  The units on the axes are arcseconds from the center of expansion of the remnant.
\normalsize }
    \label{fig:se}
\end{figure*}

We plot the Doppler components from both the 25.89$\mu$m [O~IV] and the 34.81$\mu$m [Si~II] lines in Figures \ref{fig:sw}, \ref{fig:ne}, and \ref{fig:se} for the Southwest, Northeast, and Southeast regions respectively.  We converted our velocity coordinates to spatial coordinates using the conversion factor between the two determined by \cite{delaney10}.  This conversion factor is more appropriate than one calculated from our own data since it uses data from the entire remnant rather than a few select regions.  The flux from each component is displayed by varying the transparency; the brightest 3D pixel (or "voxel") for a given ionic line is 80\% opaque, while the opacity of all other voxels is linearly scaled downwards as a function of the intensity of the Doppler component.  We have not plotted very weak ejecta with total fluxes less than 15\% that of the brightest velocity component.  The other strong line, the 33.48$\mu$m [S~III] line, is from the same nucleosynthetic layer as Si and traces out nearly identical structures to the 34.81$\mu$m [Si~II] line.  Therefore, we do not show it here.  The low density of the interior ejecta \citep[$\sim$100~cm$^{-3}$,][]{smith09} implies that self-absorption within the ejecta will be minimal along the line of sight.  We note that we are likely only observing the densest ejecta material, whether it is shocked or interior ejecta, since the emissivity should scale roughly as the density squared in both cases.

The ejecta in the SW region form a distinct shell-like structure.  The O and Si ejecta all lie along the same shell, although they fill different parts of the shell.  Averaging over the entire SW region, there are an averge of only 1.3 Doppler components per line of sight, indicating that the thickness of the shell is $\le$250 km~s$^{-1}$ along any given line of sight.  The brightest ejecta in the NE region also forms part of a shell, but there are substantially more dim ejecta inside of the shell than in the SW.  O and Si lie both on the shell and inside of the shell, although it appears that some of the O and Si is systematically separated in velocity space (see $\S$ 3.6).  The SE region consists of an irregularly shaped region of both Si and O emission in addition to a region dominated by O emission in the western most part of the region.

\subsection{Iron}

\begin{figure*}
    \centering
    \includegraphics[width=0.3\textwidth]{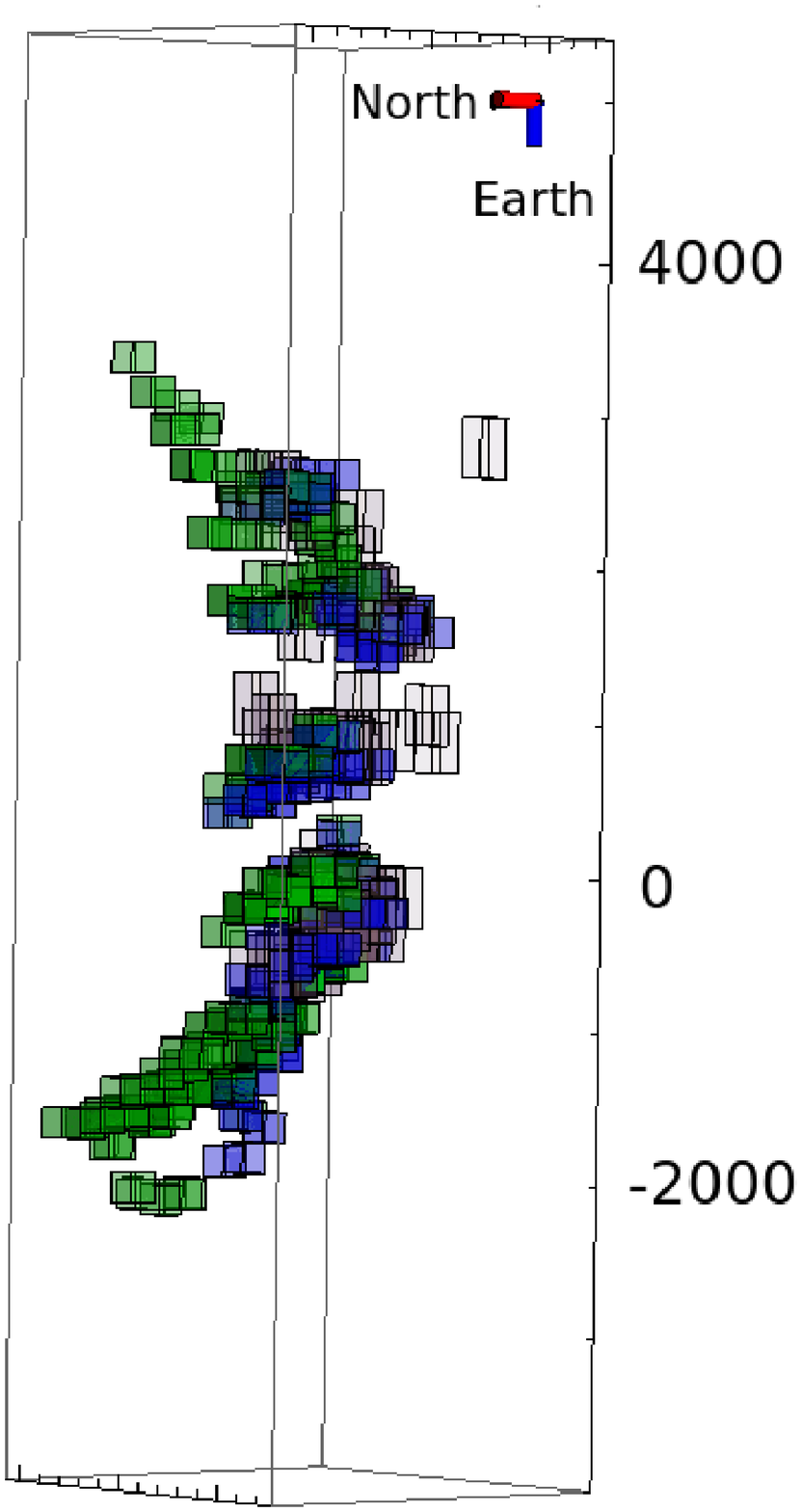}
    \includegraphics[width=0.3\textwidth]{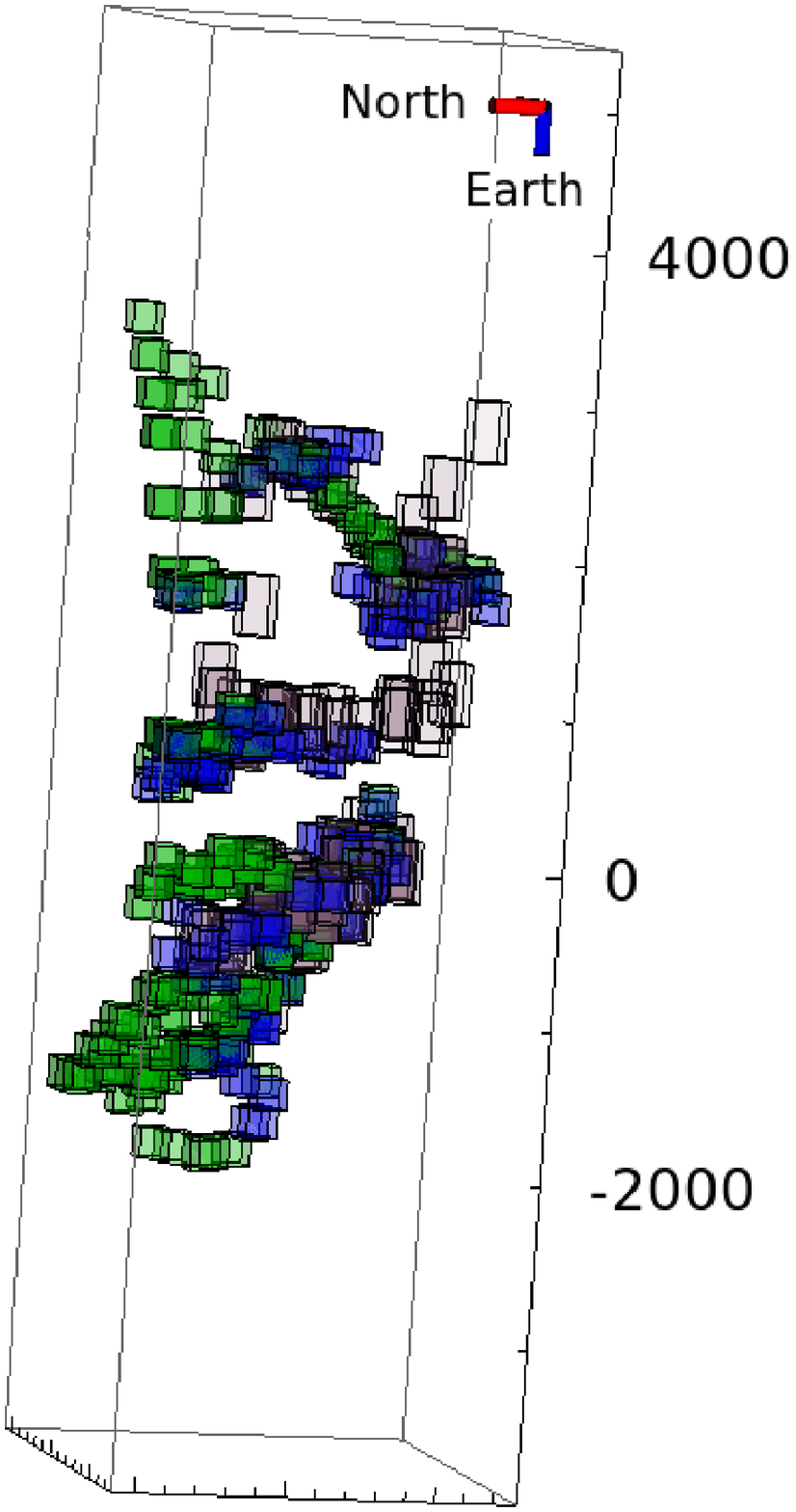}
    \includegraphics[width=0.3\textwidth]{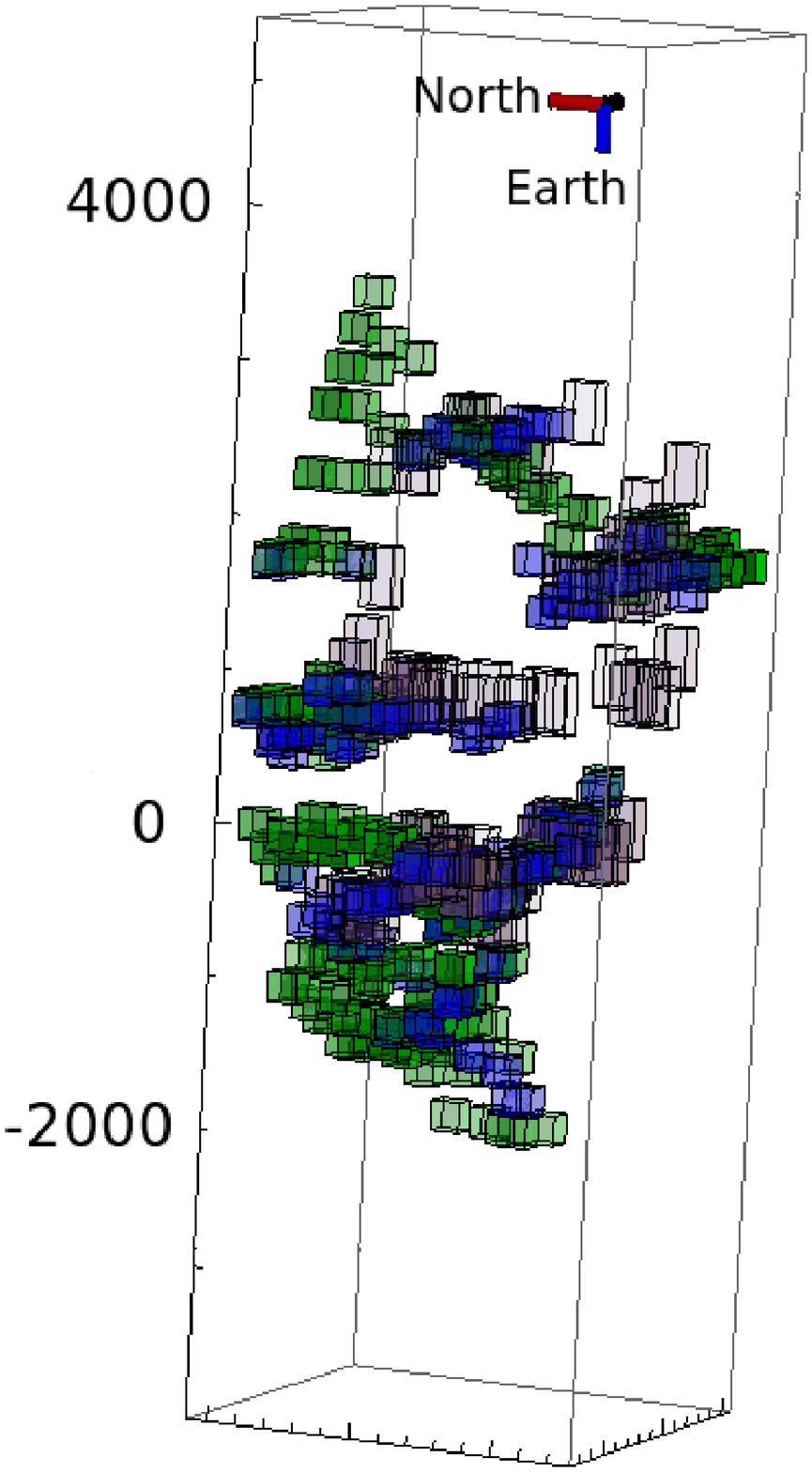}

\caption{\small 3D plot of the the 25.9$\mu$m [O~IV] line (blue), the 34.8$\mu$m [Si~II] line (green), and the 17.9$\mu$m [Fe~II] line (purple) in the Southwest of the remnant as viewed from three different angles.  The velocity axis has been stretched by a factor of approximately 1.8 in order to better highlight the Doppler structure of the region.  The Fe emission lies on the same shell as the O and Si emission.
\normalsize }
    \label{fig:femap}
\end{figure*}

We observe [Fe~II] with the SH module at 17.9$\mu$m in the SW. We plot the Doppler components from this line with Si and O emission as shown in Figure \ref{fig:femap}.  The SH data were binned 2 by 2 pixels to increase the signal to noise ratio.  It is clear that the Fe lies on the Si+O shell described in the previous section.

\begin{figure*}
    \centering
    \includegraphics[width=1.0\textwidth]{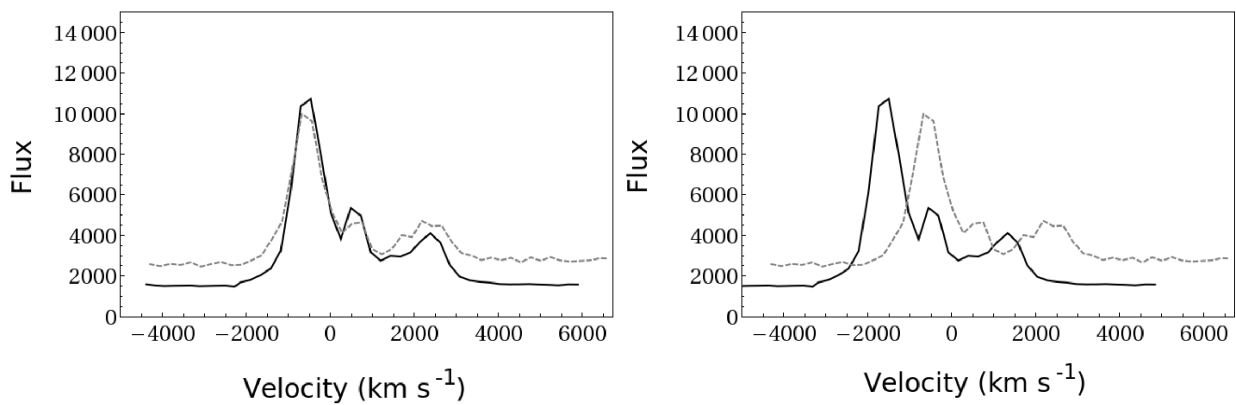}
    
\caption{\small Velocity plot for the [Si~II] line (dashed) over-plotted with the 25.9$\mu$m line (solid) shifted under the assumption that it is either all [O~IV] (left) or [Fe~II] (right).  The lines have been normalized such that the integrated flux is equal for both lines.  The velocity structure matches very well for the assumption that the 25.9$\mu$m line is all O, but matches very poorly under the assumption that it is composed of Fe.
\normalsize }
    \label{fig:dopplerO}
\end{figure*}

We confirm that the 25.9$\mu$m line is the 25.89$\mu$m [O~IV] and not the 25.98$\mu$m [Fe~II] line by comparing the Doppler structure of the 25.9$\mu$m to that of the 34.81$\mu$m [Si~II] line for several lines of sight in the SE, SW, and NE.  As an example, we display the results for one line of sight with strong 17.9$\mu$m Fe emission in Figure \ref{fig:dopplerO}.  We show the Doppler structure for the 25.9$\mu$m line under the assumption that it is all [O~IV] and all [Fe~II].  We obtain an excellent match under the assumption of [O~IV], but a poor match under the assumption of [Fe~II] even along this line of sight where we see relatively strong Fe at other wavelengths.  We find no evidence for Fe at 25.9$\mu$m for 10 other lines of sight, confirming the results of \cite{isens10}.  Therefore, we assume for the remainder of this paper that the 25.9$\mu$m line is entirely due to [O~IV] emission.

\subsection{Comparison to X-ray Emission}

\begin{figure*}
    \centering
    \includegraphics[width=0.45\textwidth]{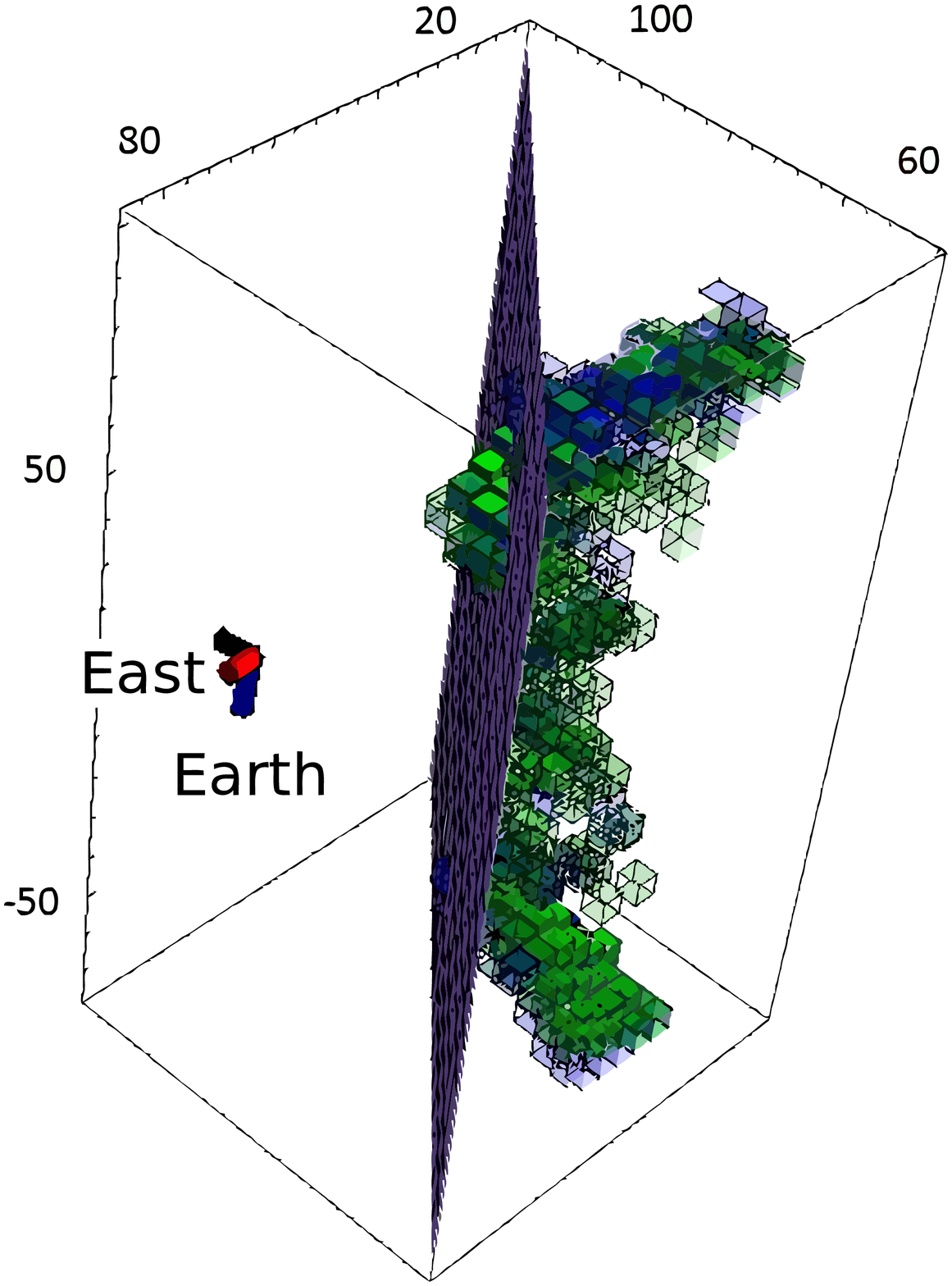}
    \includegraphics[width=0.45\textwidth]{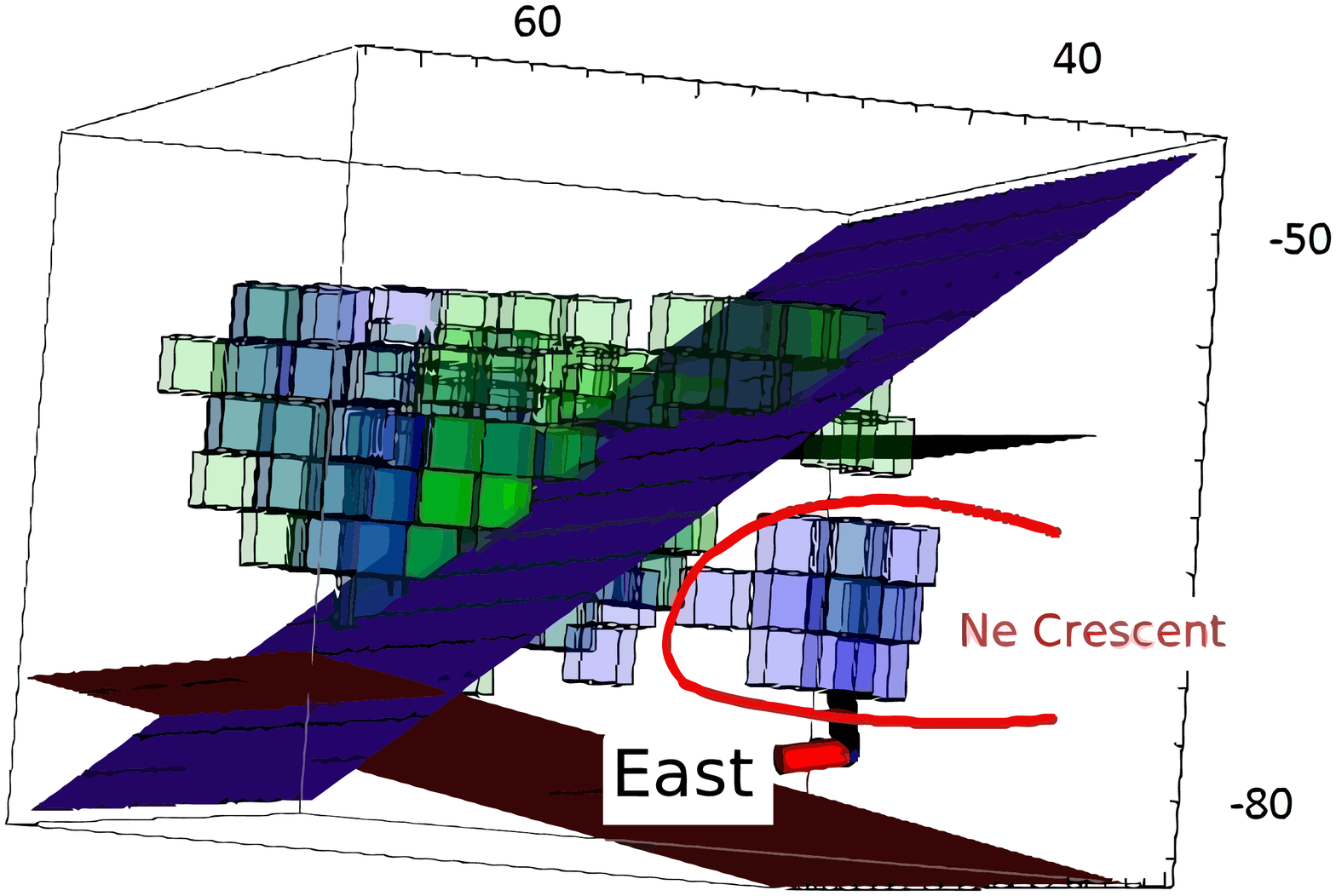}

\caption{\small 3D plot of the the 25.9$\mu$m [O~IV] line (blue) and 34.8$\mu$m [Si~II] line (green) in the Northeast (left) and Southeast (right) of the remnant along with planes of X-ray emission.  The RA and Dec of the X-ray planes were extracted from the lines on the \emph{Chandra} images in Figure \ref{fig:xrayandir}.  The IR ejecta are just behind the front edge of the X-ray shock in the Northeast.  The same is true in the Southeast, and we see a substantial amount of Oxygen dominated material where our field of view overlaps with the Ne-crescent detected by \cite{ennis06}.  This is not surprising given that Ne and O come from the same nucleosynthetic layer.
\normalsize }
    \label{fig:xrayplanes}
\end{figure*}

We compared the locations of the IR ejecta to X-ray ejecta detected in the 2004 Chandra observations of Cas~A \citep[see][]{hwang04}.  The spectral resolution of the X-ray images is not sufficient to accurately determine the Doppler velocity for most lines.  Therefore, we show the 3D location of the X-ray ejecta as planes perpendicular to the plane of the sky in Figure \ref{fig:xrayplanes}.  These planes represent the forward edge of the X-ray ejecta as seen in Figure \ref{fig:xrayandir}.  We know that the X-ray material has been recently shocked since the ejecta will only be at the appropriate ionization states and temperatures if it has recently encountered the reverse shock - the ejecta will ionize up to states that are difficult to observe in the X-ray within $\sim$100 years \citep{mazz98}.

Most of the bright IR ejecta are immediately interior to the leading edge of the X-ray material.  Since it takes some time for a plasma to up ionize to ions visible in the X-ray, this is consistent with the picture that the brightest IR ejecta have been recently shocked.

\subsection{Geometric Structure of Ejecta}

We observe that the ejecta plotted in the previous section appear to have a distinct shell-like geometry in the Southwest and Northeast regions.  We attempt to characterize the shape of this shell with an ellipsoid by fitting bright emission with a total flux at least 15\% that of the brightest Doppler components from all 3 regions.

\begin{figure*}
    \centering
    \includegraphics[width=0.7\textwidth]{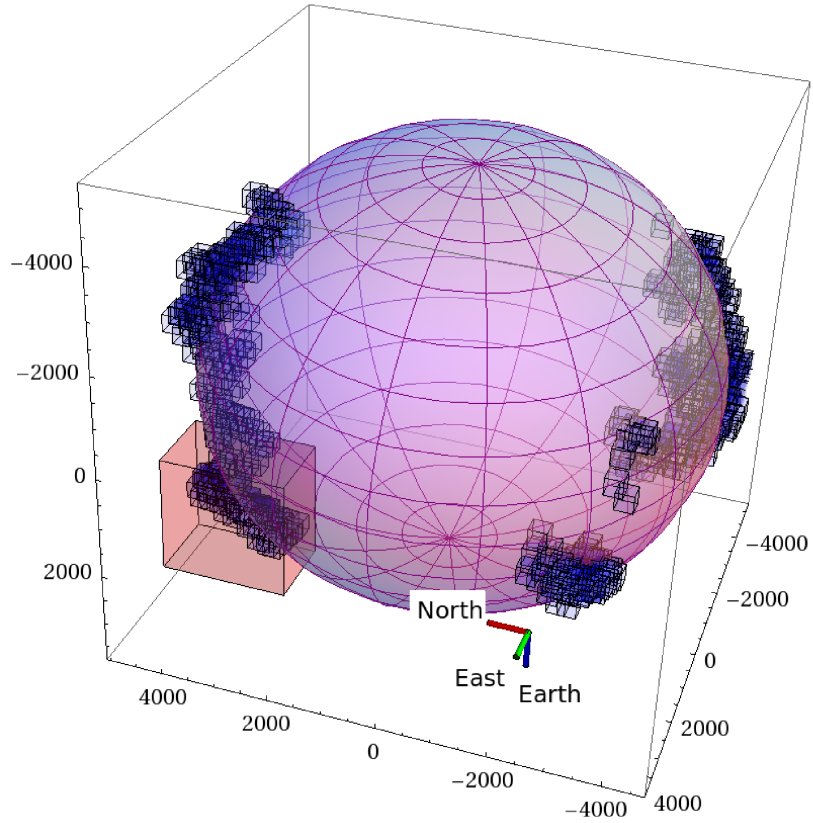}

\caption{\small 3D plot of ejecta from all regions and the best fit ellipsoid.  The units of the axes are km~s$^{-1}$.  The ellipsoid is has an eccentricity of 1.07.  The zoomed in region in Figure \ref{fig:sep} is indicated.
\normalsize }
    \label{fig:ellipsoid}
\end{figure*}

We determined the best-fit ellipsoid characterized by 8 components - the 3 spatial coordinates for the center of the ellipsoid, the 3 axes lengths, and 2 rotation angles.  We minimized the intensity weighted RMS residuals in the 3D space by iteratively stepping through all plausible combinations of parameters.  We show our best fit ellipsoid in Figure \ref{fig:ellipsoid}.  The lengths of the axes in the plane of the sky are 103\arcsec and 98.3\arcsec, and the length of the axis perpendicular to the plane of the sky is 97.2\arcsec.  The ellipsoid is a sphere to within 7\%.  The average residual from the best fit ellipsoid is 270 km~s$^{-1}$, which is roughly 5\% of the total velocity for ejecta on the ellipsoid.  The center of this sphere is offset from the geometric center of the ejecta by 810 km~s$^{-1}$ along our line of sight.  This offset can also be seen in the Velocity vs. Radius plot shown in Figure \ref{fig:vvsr}.

\subsection{Corrugation}

\begin{figure*}
    \centering
    \includegraphics[width=0.6\textwidth]{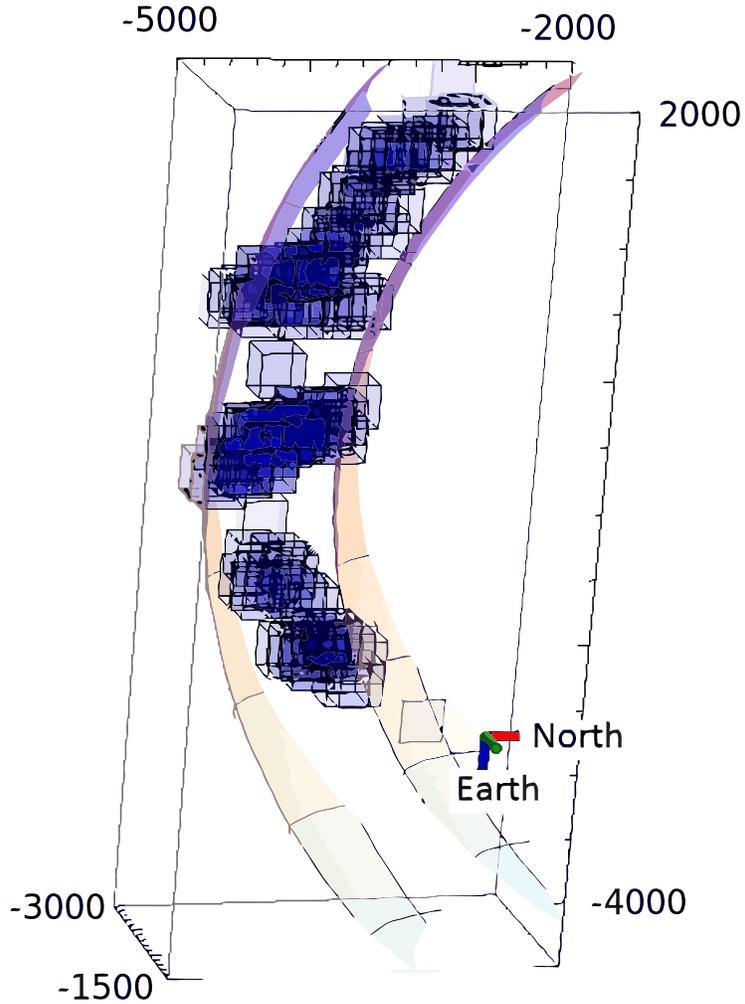}

\caption{\small 3D plot of the 25.9$\mu$m [O~IV] line in the Southwest of the remnant.  The units of the axes are km~s$^{-1}$.  Two ellipsoids with radii 250 km~s$^{-1}$ greater than and less than the best fit ellipsoid are also plotted.  Although the shell in this region is only $\sim$200 km~s$^{-1}$ thick along any single line of sight, the overall velocity of the components systematically varies by $\sim$250 km~s$^{-1}$ both above and below the best fit ellipsoid.
\normalsize }
    \label{fig:corrugation}
\end{figure*}

Although the brightest ejecta lie near a spherical surface, the ejecta appear to be corrugated about that surface.  In fact, the residuals from our best fit surface are dominated by systematic $\sim$250 km~s$^{-1}$ corrugations about the surface rather than random small-scale fluctuations.  This is most clearly seen in the Southwest region, as shown in Figure \ref{fig:corrugation}.  We find that the average wavelength of the corrugation is $\sim$24\arcsec~and the amplitude is $\sim$8\arcsec~about the best fit surface.  We further address the issue of corrugation in the next sections by looking at radial plot of the net intensity of the ejecta.

\subsection{Separation of Nucleosynthetic Layers}

\begin{figure*}
    \centering
    \includegraphics[width=0.7\textwidth]{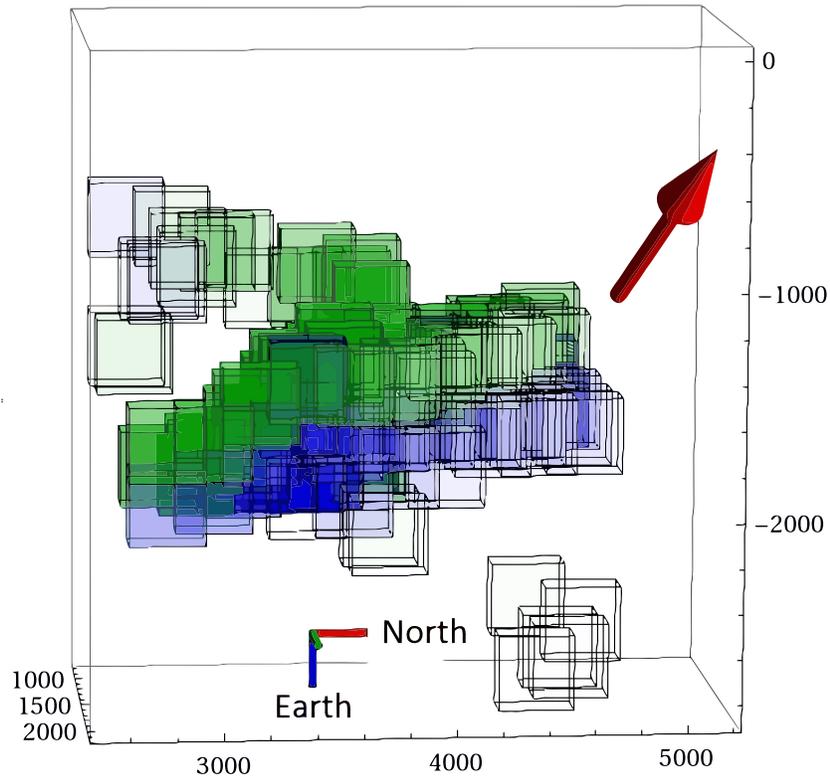}

\caption{\small 3D plot of the the 25.9$\mu$m [O~IV] line (blue) and 34.8$\mu$m [Si~II] line (green) in a select region of the Northeast whose location is indicated in Figure \ref{fig:ellipsoid}.  The units of the axes are km~s$^{-1}$.  The red arrow points to the center of the remnant.  We detect clear separation between the O and Si layers along this line of sight.
\normalsize }
    \label{fig:sep}
\end{figure*}

We plot the the brightest [Si~II] and [O~IV] ejecta, which come from different nucleosynthetic layers, for the Southwest and Northeast in Figures \ref{fig:sw} and \ref{fig:ne}.  There is clearly some separation between these layers in some directions.  We show a closeup of one such region in Figure \ref{fig:sep}, where the layers are separated by a few hundred km~s$^{-1}$ (corresponding to $\sim$5\arcsec or $\sim$0.1 pc).  The location of this region is shown in Figure \ref{fig:ellipsoid}.

In the SE, we observe a clump of O rich ejecta with almost no corresponding Si emission.  We address this interesting region in more depth in $\S$4.2.

We further examine and quantify the separation between layers by plotting the intensity of the emission as a function of three-dimensional radius for all intensities of ejecta for both O and Si in the Southwest.  Since Doppler velocity and spatial coordinates are equivalent and we want to determine the velocity separation between nucleosynthetic layers, we converted all our spatial coordinates into velocity units in order to determine the 3 dimensional velocity from the center of expansion found in $\S$3.3.  This was accomplished by using the arcseconds to km~s$^{-1}$ ratio determined by \cite{delaney10}.  We then plotted the line flux as a function of 3D velocity by binning the emission in 200 km~s$^{-1}$ increments.  

Because we expect the behavior of the layers to vary as a function of direction, we created plots for many different lines of sight.  For each plot we only plotted emission from a solid angle $\pi$/12 steradians wide.  Our initial beam was centered $\pi$/8 radians above the plane of the sky and is wide enough to include the entire width of the region, and then incremented by $\pi$/12 radians downward for subsequent lines of sight.

\begin{figure*}
    \centering
    \includegraphics[width=0.9\textwidth]{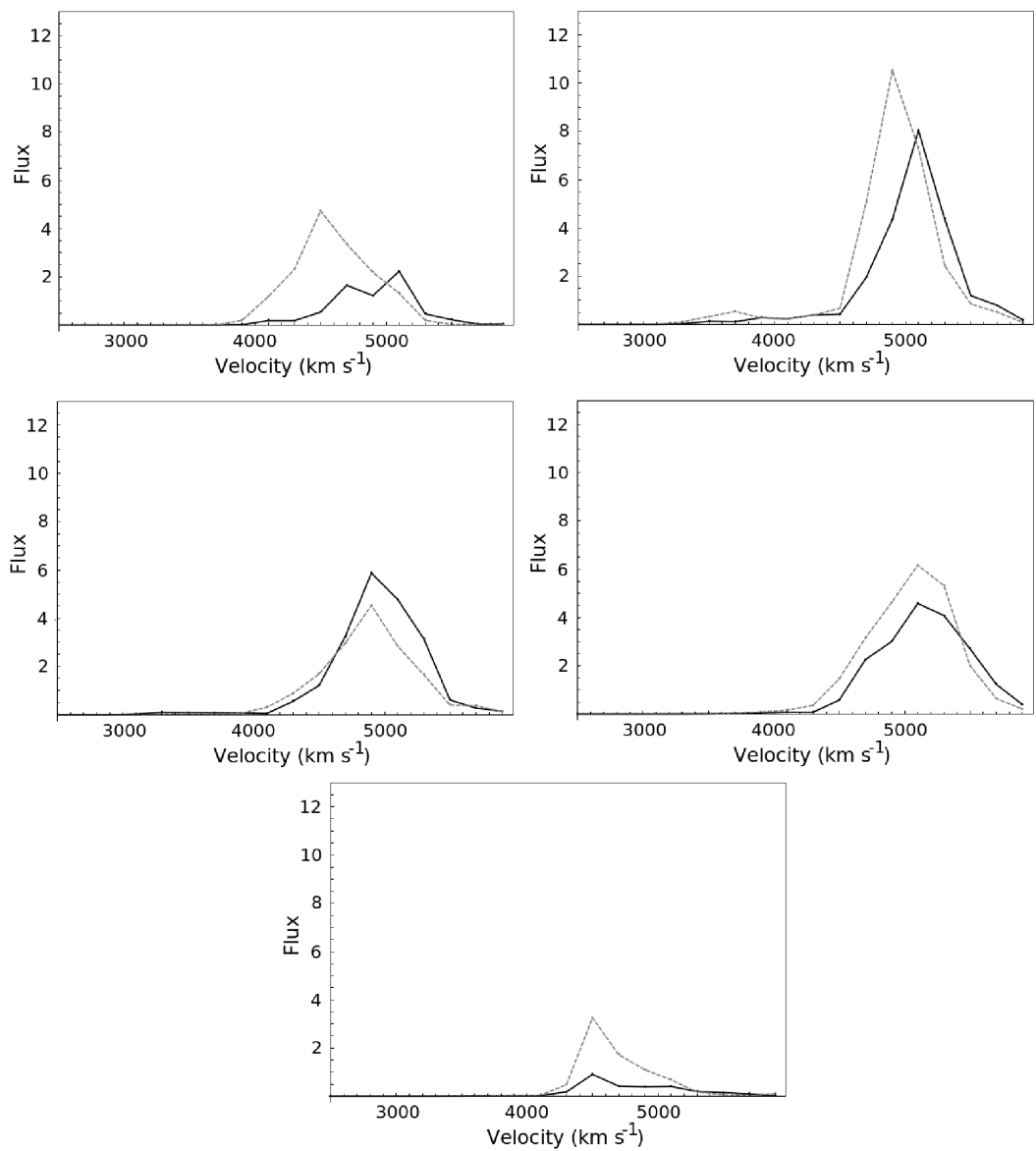}

\caption{\small Flux vs radius plot for the Southwest region.  The O (dashed) and Si (solid) distributions overlap along some lines of sight, but are offset by up to $\sim$500 km s$^{-1}$ along others.  The peak velocity of the O line and the Si line also vary from about 4400 km s$^{-1}$ to 5200 km s$^{-1}$ depending on the line of sight.  This is a clear signature of corrugation.  The full-width-half-max of the distributions is $\sim$1000 km s$^{-1}$ along all lines of sight.
\normalsize }
    \label{fig:ivsr}
\end{figure*}

We show the flux vs. 3D velocity plots for the Southwest in Figure \ref{fig:ivsr}.  The average radial distance of the O with respect to Si varies significantly between locations.  Along some lines of sight, they overlap to within one 2\arcsec~bin.  In different lines of sight, the O peak is at a velocity up to $\sim$500 km~s$^{-1}$ greater than the Si.  And along yet other lines of sight, the Si and O peak at roughly the same velocity, but much of the O still is at larger radii than the Si.

Furthermore, the peak velocity of both the Si and O changes as a function of direction from $\sim$4400 km~s$^{-1}$ to $\sim$5200 km~s$^{-1}$.  This is consistent with corrugation in the velocities of the ejecta along different directions.

The velocity width averaged over the solid angle of the Si and O ejecta in the previous section are physical widths, and not just instrumental effects.  We find that, to first order, the FWHMs are all $\sim$1000 km~s$^{-1}$ for both Si and O.  The velocity uncertainty in our bins is roughly 130 km s$^{-1}$ (since we have an uncertainty of roughly 65 km~s$^{-1}$ in each direction) and the bins are 200 km~s$^{-1}$ wide.

We note that the velocity width of the peaks is much larger than the velocity width for any given line of sight determined in $\S$3.1.  The velocity widths of the peaks is dominated by averaging the 3D velocities over the entire solid angle and is not necessarily indicative of the velocity spread over any single line of sight.

\subsection{Faint Ejecta}

\begin{figure*}
    \centering
    \includegraphics[width=0.45\textwidth]{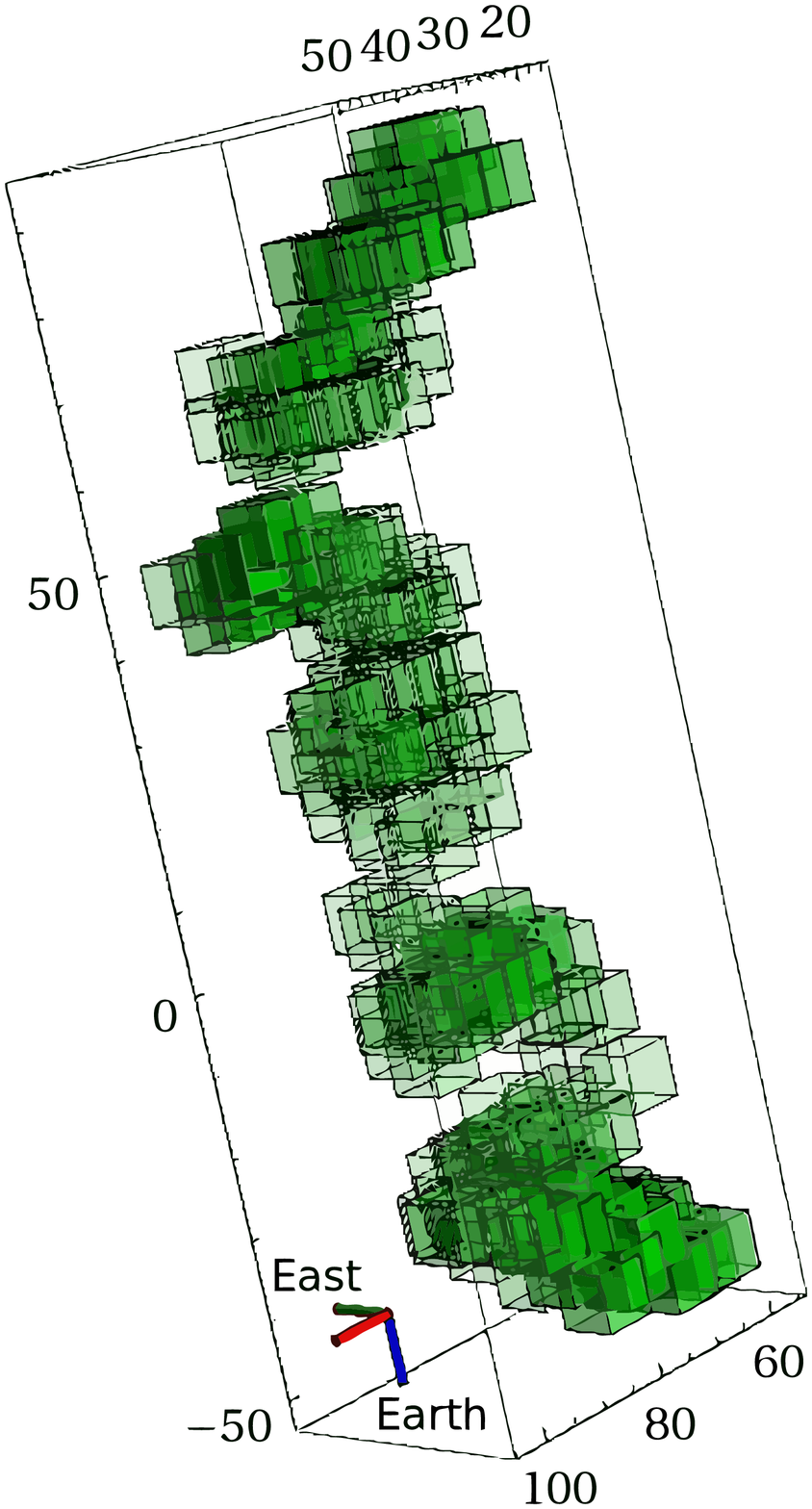}
    \includegraphics[width=0.45\textwidth]{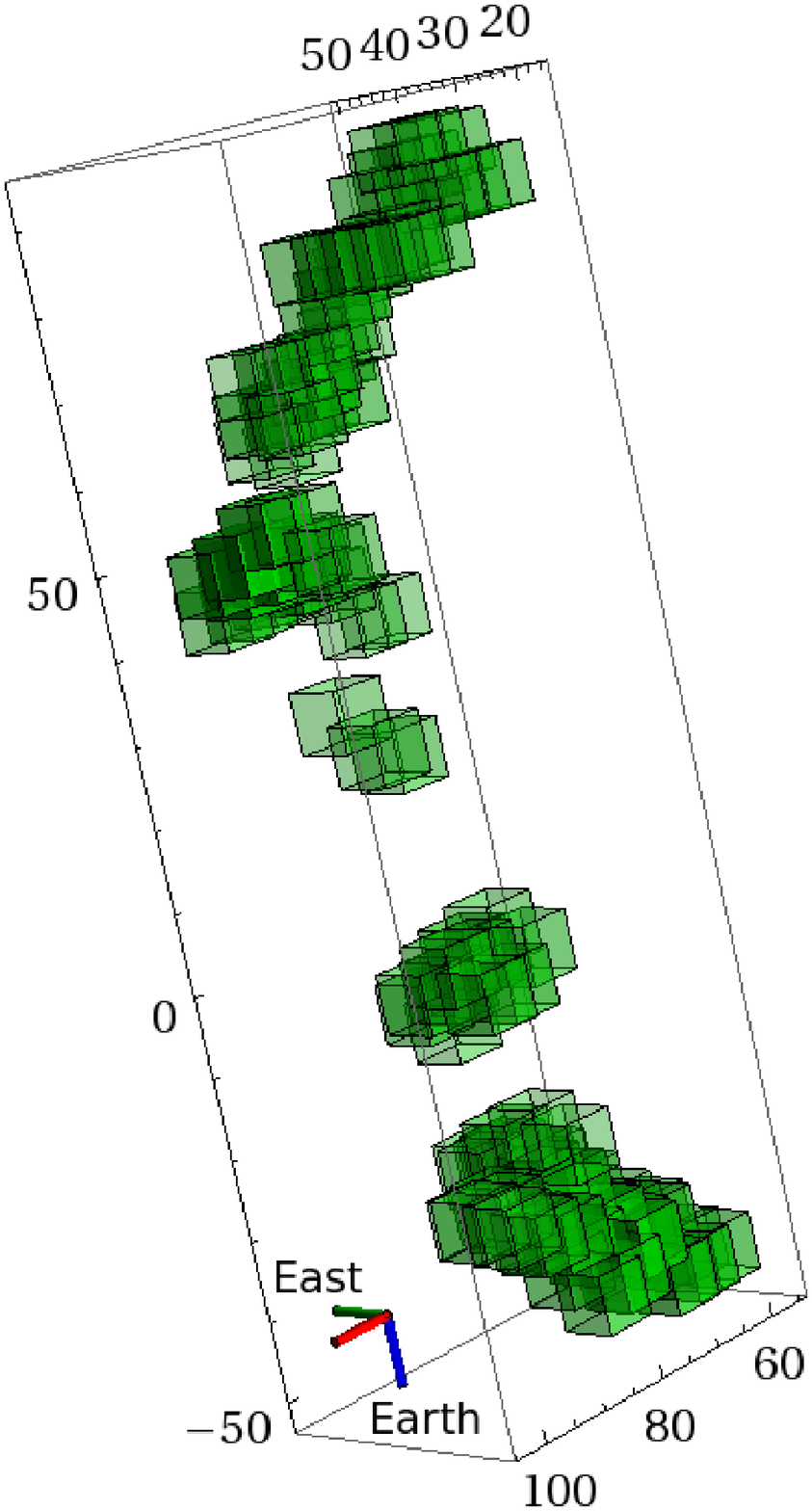}

\caption{\small 3D plot of all ejecta in the Northeast region with at least 10\% the flux of the brightest Doppler component (left) and all ejecta with at least 25\% the flux of the brightest Doppler component (right) .  The units on the axes are arcseconds from the center of expansion of the remnant.  The dimmer ejecta lie interior to the bright shell formed by the brightest ejecta, indicating that it may not yet have encountered the reverse shock.
\normalsize }
    \label{fig:dim}
\end{figure*}

Although we dealt mostly with bright ejecta that are found to lie on a distinct spherical structure in the previous sections, there are weaker ejecta which lie interior to this bright shell, especially in the NE region.  On average, these ejecta are $\sim$10\% as bright as the ejecta on the shock.  We plot these ejecta along with the bright material in Figure \ref{fig:dim}.  We note that nearly all the dim ejecta lie interior to the bright ejecta.

\section{Discussion}

\subsection{Supernova Explosion Physics}

The nature of core-collapse supernova explosions is a major area of research.  The assumed structure of the star before the supernova explosion is similar for many different models.  As a massive star fuses different elements during hydrostatic burning, it should produce denser and denser concentric nucleosynthesis layers, forming the classic ``onion-skin'' model of the star.

However, between models, there is substantial variation in the relevant physics behind the supernova explosion itself.  Most groups propose neutrino-driven shocks as the main mechanism causing the explosion, but some utilize diffusive, magnetic buoyancy, or neutrino-bubble instabilities \citep{janka07}.  Other groups propose jet driven explosions, where the explosion is dominated by MHD driven jets formed in rapidly rotating stars \citep[e.g.][]{burrows07}.

3D maps of different nucleosynthetic layers in Cas~A provide a unique opportunity to test and constrain the various models.  Our observations focus on the layers that were initially near the core of the star, where the supernova explosion begins - the Fe/Ni core, the Si/S layer immediately above the core, and the O/Ne layer above the Si/S.  We can observe the post-explosion geometry and velocity profile of the different layers.  Some models predict that Si and O will be ejected at nearly the same velocity \citep[e.g.][]{kifonidis06}, while others predict that they will be ejected at velocities that differ by $\ge$500 km s$^{-1}$ \citep[e.g.][]{jog09}.

Along many lines of sight, our results are consistent with the models of \cite{kifonidis06} and the 25M$_{\odot}$ models of \cite{jog09} - we see little difference between the peak velocities of O and Si.  However, along other lines of sight, we find that the O and Si peaks are offset by hundreds of km~s$^{-1}$ which is inconsistent with above models, but is consistent with the results of the 15M$_{\odot}$ models of \cite{jog09}.  However, even though this 15M$_{\odot}$ model produces a reasonable separation between O and Si, it predicts that the velocity width of O will be nearly twice that of Si, while we observe that both layers have roughly the same width along all lines of sight.  Furthermore, the model predicts overall velocities of $<$2000 km~s$^{-1}$, which is only half of what we observe.  Put together, available models can reproduce most of the various behaviors that we observe, but no single model, so far, can reproduce observed velocity structure of the nucleosynthetic layers of the Cas~A supernova explosion.

\subsection{The Southeast}

\cite{ennis06} found a Neon ``crescent'' of ejecta in the Southeast and Northeast of Cas~A where Neon was seen in the IR, but little Si was detected in the IR or X-ray.  Our Southeast region overlaps slightly with the Neon crescent.  Most of the IR emission in this region consists of overlapping Si and O immediately behind the X-ray ejecta.  However, in the Neon crescent, we see only O emission that is more than an order of magnitude stronger than Si emission as shown in Figure \ref{fig:xrayplanes}.  This is consistent with \cite{ennis06} and \cite{smith09} since Neon and O come from the same nucleosynthetic layer.  The most likely explanation for this Neon/O crescent is that the Neon and O are currently encountering the reverse shock and therefore becoming brighter as they are compressed, while the Si has yet to encounter the reverse shock.  This is consistent with the picture that there is no IR emission from the Si because the densest Si clumps have not yet been radiatively shocked, and there is no observable X-ray emission because the Si ejecta have not been non-radiatively shocked.

\subsection{Geometry}

We observe that the center of expansion is offset from the center of the remnant by $\sim$810 km/s along our line of sight.  This is consistent with previous results from 3D reconstructions in the optical \citep{reed95} and the IR \citep{delaney10, isens10}.  There are three major possible sources for the cause of this offset - asymmetries in the circumstellar environment, movement by the progenitor star, and the supernova explosion itself.

\cite{reed95} speculated that this offset was due to asymmetries in the pre-supernova circumstellar environment.  \cite{isens10}, using one patch toward the center of the remnant, argue that this cannot be the case since the ejecta interior to the reverse shock show the same velocity asymmetry, despite the fact that they are unaffected by the circumstellar material.  The interior ejecta are expanding into a bubble that has been cleared of any circumstellar material by the shocks associated with the supernova explosion.  Since our new results, which span the Bright Ring, are consistent with those of \cite{isens10}, the velocity offset is not specific to the central regions of Cas~A, but applies to the entire remnant.

In principle, the velocity offset could be caused by progenitor motion, but the observed 810 km~s$^{-1}$ offset is much too large a velocity for a star.  Neutron stars may have velocities $>$500 km~s$^{-1}$ \citep[e.g.][]{delaney11}, but such large velocities are the result of ``kicks'' during the supernova explosion, which are caused by asymmetries in the supernova explosion itself or a binary companion \citep[e.g.][]{sh69, chat05}.

Therefore, the observed offset it most likely caused by asymmetries in the supernova itself.  A likely culprit is asymmetries formed in the first $\sim$100 milliseconds as seen in the models of \cite{burrows07} and the SASI models of \cite{blondin03}.  Both of these instabilities allow the initially spherically symmetric forward shock to become highly asymmetric in just a few crossing times.  These instabilities arise due to the response of the post shock pressure to changes in the shock radius.  If the pressure in one region becomes slightly higher than the surroundings, it will push the shock outward.  The preshock pressure drops with increasing radius, which leads to smaller pressures behind the forward shock due to the outward shock displacement.  If the postshock pressure radial profile is steeper than the preshock pressure profile, a standing acoustic wave is produced by the positive feedback loop.  Ejecta can ``slosh'' between the standing shocks, resulting in substantial asymmetries \citep{blondin03}.  Presumably, the ejecta will maintain this asymmetry as they expand outward, resulting in a low-order asymmetry that is not necessarily centered on the location of the progenitor star.  Note that although the shock may initially be asymmetric, it will gradually become spherical over time \citep[e.g.][]{bb82}, producing the nearly spherical surface that we observe today.

\subsection{Faint Ejecta}

The faint ejecta seen interior to the bright ejecta on the nearly spherical shell are likely ejecta that have yet to encounter the reverse shock.  We expect to observe both [Si~II] and [O~IV] even if they are not yet shocked because they will be photoionized by energetic UV and X-ray photons from the reverse shock.  Such ejecta were previously detected in the center of the remnant and interior to the reverse shock by \cite{delaney10}, \cite{smith09}, and \cite{isens10}.  The brightness of the ejecta appears to be roughly an order of magnitude less than that of nearby material which has been shocked.  This is consistent with what is expected from a strong shock - a compression factor of about 4 is expected for a classic, strong, non-radiative shock, which would cause a rise in emissivity of a factor of 16.

We note that the interior ejecta in the center of the remnant (see Figure \ref{fig:xrayandir}) discussed extensively in \cite{isens10} are much brighter than the interior ejecta in all the regions near the reverse shock.  The most obvious explanation for this is that the central ejecta are at a higher density.  This difference in density is probably caused by geometric effects - the central material from the remnant is traveling at about half the velocity of the material currently encountering the reverse shock in the plane of the sky \citep{delaney10}.  If the ejecta are expanding homologously and were initially ejected at approximately the same density, the material in the place of the sky would be at a density one quarter that of the interior ejecta since the shocked ejecta are at twice the radius.  Therefore, we would expect the emissivity of the central ejecta to be roughly 16 times that of material in the plane of the sky that is interior to the reverse shock since the emissivity of the ejecta varies as the square of the density.  This is roughly what we observe - interior material in these regions have an average brightness of $\sim$300 MJy sr$^{-1}$, while material in the center of the remnant have brightnesses around 4000 MJy sr$^{-1}$ \citep{isens10}.


\subsection{Corrugation}

Corrugation - that is, ripples in the geometric structure of the ejecta -  has been previous observed in several supernova remnants, including SN1006 \citep[e.g.][]{wl97} and the Cygnus Loop \citep[e.g.][]{ray03}.  There are several possible explanations for this corrugation.  If the shock is radiative, the ripples could be caused by the thermal instability \citep[e.g.][]{bert86} or the thin shell instability \citep{vish83}.  The thermal instability is especially relevant for high speed ($>$150 km s$^{-1}$), high temperature (T$>$10$^{5}$ K) shocks where the sound speed crossing time greatly exceeds the cooling time.  This ``instability'' is actually an overstability that results from the high radiative cooling rate.  The thin shell instability is another overstability.  In this scenario, the ram pressure and thermal pressure are misaligned, causing ripples in the initially smooth distribution of ejecta.


Another possibility is that the ripples are caused by inhomogeneities in the ejecta encountering the shock.  This inhomogeneity could be caused by Rayleigh-Taylor filaments created at the contact discontinuity between the ejecta and the ISM as the remnant enters its Sedov-Taylor phase \citep{wc01}.  While this is a likely explanation for the corrugation observed in SN1006 \citep{long03}, it is not likely in Cas~A since the ISM has already been swept away by the forward shock by time the ejecta encounter the reverse shock.

A final explanation is that the corrugation is caused by density variations in the ejecta itself \citep[e.g.][]{ray03}.  In this model, dense ejecta encountering the reverse shock would be slowed less than less dense clumps of ejecta.  \cite{ray03} argues that this instability is the most likely cause for the corrugation in the Cygnus Loop, where the density variations may have been caused by ISM turbulence.


Regardless of the mechanism, there is also a question of where the corrugation occurs.  The observed ejecta have encountered several shocks - the forward shock during the supernova explosion itself, a reverse shock during the supernova, and most recently, a larger scale reverse shock in the supernova remnant \citep{isens10}.  Which shock encounter creates the observed corrugation?

Most studies expect corrugation at the second, larger forward or reverse shocks associated with the remnant.  However, we conclude that the most likely location for the corrugation is during one of the two shocks that the ejecta encounter during the supernova explosion itself.  Previous studies find little evidence for deceleration of IR ejecta after the explosion, including recently shocked ejecta \citep{delaney10}.  Our data are also inconsistent with recent deceleration of the ejecta.  If the ejecta were suddenly decelerated, the ejecta would be closer to the plane of they sky in our models due to the reduction in Doppler velocity, forming a flattened surface.  However, we do not observe this pattern in the shocked ejecta - we still see a nearly perfect sphere.  In other words, if the corrugation were being caused at the reverse shock, the ejecta would need to be decelerated by at least several hundred kilometers per second in order to create the observed corrugation since being shocked.  This effect would be easily visible with our Doppler reconstruction of the global geometry, but is not observed.  Thus, the corrugation must occur during the supernova explosion itself, perhaps at the initial forward or explosion reverse shock.

\section{Conclusions}

We create a 3D model of shocked ejecta of Cas~A in select regions at unprecedented spectral resolution using IR ionic lines.  We confirm previous studies that indicate that the remnant is offset by $\sim$800 km~s$^{-1}$ along our line of sight.  We find evidence for velocity separation between the O and Si layers along some, but not all, lines of sight.  We measure the velocity width of these layers roughly 250 km~s$^{-1}$ thick for a single line of sight, although the ejecta are often in bands that are $\sim$1000 km~s$^{-1}$ thick averaged over several nearby lines of sight due to corrugation.  We find evidence for corrugation in some regions of the remnant, and speculate that the corrugation was caused during the explosion itself rather than hundreds of years later.  We use our observations of Si and O velocities to begin constraining models of supernova explosions, and to motivate future models to explore velocity profiles as a function of azimuth.

We look forward to potential similar data sets from instruments such as the Herschel Space Observatory and the Stratospheric Observatory for Infrared Astronomy (SOFIA).  Both these observatories will have the ability to create spectral cubes of Cas~A at much higher spectral resolution.  The current instruments on both observatories do not have the necessary instantaneous bandwidth to observe the 10,000 km~s$^{-1}$ velocity range of ejecta in Cas~A, but future spectrographs will have scanning modes that will allow observations of ejecta at many different velocities.

\acknowledgements

This work is based on observations made with the Spitzer Space Telescope, which is operated by the Jet Propulsion Laboratory, California Institute of Technology under NASA contract 1407.  This work was supported in part by NASA/SAO Award No. AR5-6008X and NASA/JPL through award 1265552 to the University of Minnesota.

K.~I. would like to thank Alexander Heger for valuable insight into the physics of supernova explosions.

\newpage

	
\begin{thebibliography}{}


\bibitem[Berschinger (1986)]{bert86} Bertschinger, E. 1986, ApJ, 304, 154
\bibitem[Bisnovatyi-Kogan \& Blinnikov (1982)]{bb82} Bisnovatyi-Kogan, G. S., \& Blinnikov, S. I. 1982, SvA, 26, 530
\bibitem[Blair et al. (2000)]{blair00} Blair, W. P., Morse, J. A., Raymond, J. C., Kirshner, R. P., Hughes, J. P., Dopita, M. A., Sutherland, R. S., Long, K. S., \& Winkler, P. F. 2000, ApJ, 537, 667
\bibitem[Blondin et al. (2003)]{blondin03} Blondin, J.M., Mezzacappa, A., \& DeMarino, C. 2003, ApJ, 584, 971
\bibitem[Braun (1987)]{braun87} Braun, R. 1987, A\&A, 171, 233
\bibitem[Burrows et al. (2007)]{burrows07} Burrows, A., Dessart, L., Ott, C. D., \& Livne, E. 2007, PhR, 442, 23
\bibitem[Chatterjee et al. (2005)]{chat05} Chatterjee, S. et al. 2005, ApJ, 630, L61
\bibitem[Chakrabarty et al. (2001)]{chak01} Chakrabarty, D., Pivovaroff, M.J., Hernquist, L.E., Heyl, J.S., \& Narayan, R. 2001, ApJ, 548, 800
\bibitem[Chevalier \& Kirshner (1978)]{chev78} Chevalier, R. A., \& Kirshner, R. P. 1978, ApJ, 219, 931
\bibitem[Chevalier \& Kirshner (1979)]{chev79} Chevalier, R. A., \& Kirshner, R. P. 1979, ApJ, 233, 154
\bibitem[Chevalier \& Oishi (2003)]{chev03} Chevalier, R. A., \& Oishi, J. 2003, ApJ, 593L, 23
\bibitem[Chevalier \& Soker (1989)]{chev89} Chevalier, R. A., \& Soker, N. 1989, ApJ, 341, 867
\bibitem[DeLaney et al.(2010)]{delaney10} DeLaney, T.A., Rudnick, L., Stage, M.D., Smith, J.D., Isensee, K.I., Rho, J., Allen, G.E., Gomez, H., Kozasa, T., Reach, W.T., Davis, J.E., \& Houck, J.C.  2010, in preparation
\bibitem[Ding et al. (1999)]{ding99} Ding, Y., Li, Z., \& Diwu, Y. 1999, ChA\&A, 23, 484
\bibitem[Douvian et al. (1999)]{douv99} Douvion, T., Lagage, P. O., \& Cesarsky, C. J. 1999, A\&A, 352, L111
\bibitem[Ennis et al.(2006)]{ennis06} Ennis, J. A., Rudnick, L., Reach, W. T., Smith, J. D., Rho, J., DeLaney, T., Gomez, H., \& Kozasa, T. 2006, ApJ, 652, 376
\bibitem[Eriksen et al.(2009)]{ericksen09} Eriksen, K. A., Arnett, D., McCarthy, D.W., \& Young, P. 2009, ApJ, 697, 29
\bibitem[Feuchtgruber et al.(1997)]{feucht97} Feuchtgruber, H. et al. 1997, ApJ, 487, 962
\bibitem[Fesen \& Gunderson(1996)]{fg96} Fesen, R. A. \& Gunderson, K. S. 1996, ApJ, 470, 967
\bibitem[Fesen et al.(2006)]{fesen06} Fesen, R. A., Hammell, M. C.,Morse, J., Chevalier, R. A., Borkowski, K. J., Dopita, M. A., Gerardy, C. L., Lawrence, S. S., Raymond, J. C., \& van den Bergh, S. 2006, ApJ, 645, 283
\bibitem[Fesen et al.(2001)]{fesen01} Fesen, R. A., Morse, J. A., Chevalier, R. A., Borkowski, K. J., Gerardy, C. L., Lawrence, S. S., \& van den Bergh, S. 2001, AJ, 122, 2644
\bibitem[Froese Fischer (1983)]{ff83} Froese Fischer, C. 1983, J. Phys. B, 16, 157
\bibitem[Hamilton \& Sarazin(1984)]{hs84} Hamilton, A. J. S. \& Sarazin, C. L. 1984, ApJ, 287, 282
\bibitem[Hamilton \& Fesen(1998)]{hf98} Hamilton, A. J. S. \& Fesen, R. A. 1998, ApJ, 327, 178
\bibitem[Hammer et al. (2010)]{hammer10} Hammer, N. J., Janka, H.-Th., \& Müller, E. 2010, ApJ, 714, 1371
\bibitem[Herant \& Woosley (1994)]{hw94} Herant, M., \& Woosley, S.E. 1994, ApJ, 425, 814
\bibitem[Hughes et al. (2000)]{hughes00} Hughes, J. P., Rakowski, C. E., Burrows, D. N., \& Slane, P. O. 2000, ApJ, 528, L109
\bibitem[Hwang \& Laming (2003)]{hwl03} Hwang, U., \& Laming, J. M. 2003, ApJ, 597, 362
\bibitem[Hwang et al. (2004)]{hwang04} Hwang, U., et al. 2004, ApJ, 615, L117
\bibitem[Isensee et al. (2010)]{isens10} Isensee, K., Rudnick, L., DeLaney, T.A., Smith, J. D., Rho, J., Reach, W. T., Kozasa, T., \& Gomez, H. 2010, ApJ, 725, 2059
\bibitem[Janka et al. (2007)]{janka07} Janka, H.-Th., Langanke, K., Marek, A., Martínez-Pinedo, G., \& Müller, B. 2007, PhR, 442,38
\bibitem[Joggerst et al. (2009)]{jog09} Joggerst, C.C., Woosley, S.E., \& Heger, A. 2009, ApJ, 693, 1780
\bibitem[Kassim et al.(1995)]{kas95} Kassim, N. E., Perley, R. A., Dwarakanath, K. S., \& Erickson, W. C. 1995, ApJ, 455, L59
\bibitem[Khokhlov et al. (1999)]{kho99} Khokhlov, A. M., Höflich, P. A., Oran, E. S., Wheeler, J. C., Wang, L., \& Chtchelkanova, A. Yu. 1999, ApJ, 524L, 107
\bibitem[Kifonidis et al. (2006)]{kifonidis06} Kidonidis, K., et al. 2006, A\&A, 453, 661
\bibitem[Krause et al. (2008)]{krause08} Krause, O., Birkmann, S.M., Usuda, T., Hattori, T., Goto, M., Rieke, G. H., \& Misselt, K. A. 2008, Science, 320, 1195
\bibitem[Laming \& Hwang (2003)]{lhw03} Laming, J. M., \& Hwang, U. 2003, ApJ, 597, 347
\bibitem[Laming et al. (2006)]{laming06} Laming, J. M., Hwang, U., Radics, B., Lekli, G., \& Takács, E. 2006, ApJ, 644, 260
\bibitem[Lawrence et al.(1995)]{law95} Lawrence, S. S., MacAlpine, G. M., Uomoto, A., Woodgate, B. E., Brown, L. W., Oliverson, R. J., Lowenthal, J. D., \& Liu, C. 1995, AJ, 109, 2635
\bibitem[Long et al. (2003)]{long03} Long, K. S., Reynolds, S. P., Raymond, J. C., Winkler, P. F., Dyer, K. K., \& Petre, R. 2003, ApJ, 586, 1162
\bibitem[Mazzotta et al. (1998)]{mazz98} Mazzotta, P., Mazzitelli, G., Colafrancesco, S., \& Vittorio, N. 1998, A\&AS, 133, 403
\bibitem[Miles (2009)]{miles09} Miles, A.R. 2009, ApJ, 696, 498
\bibitem[Morse et al.(2004)]{morse04} Morse, J. A., Fesen, R. A., Chevalier, R. A., Borkowski, K. J., Gerardy, C. L., Lawrence, S. S., \& van den Bergh, S. 2004, ApJ, 614, 727
\bibitem[Osterbrock \& Ferland (2006)]{ost06} Osterbrock, D. E. \& Ferland, G. J. 2006, Astrophysics of Gaseous Nebulae and Active Galactic Nuclei, 2nd. ed. (Sausalito, CA: Univ. Sci. Books)
\bibitem[Patnaude \& Fesen (2007)]{pf07} Patnaude, D.J., \& Fesen, R.A. 2007, AJ, 133, 147
\bibitem[Reed et al.(1995)]{reed95} Reed, J. E., Hester, J. J., Fabian, A. C., \& Winkler, P. F. 1995, ApJ, 440, 706
\bibitem[Raymond (2003)]{ray03} Raymond, J.C. 2003, RevMexAA, 15, 258 
\bibitem[Reynolds et al.(2008)]{rey08} Reynolds, S. P., Borkowski, Kazimierz K.J., Green, D. A., Hwang, U., Harrus, I., Petre, R.A. 2008, ApJ, 680L, 41
\bibitem[Rho et al.(2008)]{rho08} Rho, J., Kozasa, T., Reach, W. T., Smith, J. D., Rudnick, L., DeLaney, T., Ennis, J. A., Gomez, H., Tappe, A. 2009, ApJ, 673, 271
\bibitem[Satterfield et al. (2011)]{delaney11} Satterfield, J., DeLaney, T., Chatterjee, S. 2011, AAS Poster, 21725615S
\bibitem[Shklovskii (1969)]{sh69} Shklovskii, I. S. 1969, AZh, 46, 715
\bibitem[Smith et al.(2007)]{smith07} Smith, J. D., Armus, L., Dale, D. A., Roussel, H., Sheth, K., Buckalew, B. A., Jarrett, T. H., Helou, G., Kennicutt, R. C., Jr. 2007, PASP, 119, 1133
\bibitem[Smith et al.(2009)]{smith09} Smith, J. D., Rudnick, L., DeLaney, T.A., Rho, J., Gomez, H., Kozasa, T., Reach, W.T., Isensee, K.A 2009, ApJ, 693, 713
\bibitem[Tananbaum(1999)]{tan99} Tananbaum, H. 1999, IAU Circ. 7246
\bibitem[Thorstensen et al.(2001)]{thor01} Thorstensen, J. R., Fesen, R. A., \& van den Bergh, S. 2001, AJ, 122, 297
\bibitem[Vink et al.(1996)]{vink96} Vink, J., Kaastra, J. S., Bleeker, J. A. M. 1996, A\&A, 307L, 41
\bibitem[Vishniac (1983)]{vish83} Vishniac, E. T. 1983, ApJ, 274, 152
\bibitem[Wang \& Chevalier (2001)]{wc01} Wang, L., \& Chevalier, R.A. 2001, ApJ, 549, 1119
\bibitem[Wang \& Wheeler(2008)]{wang08} Wang, L., \& Wheeler, J. C. 2008, ARA\&A, 46, 433
\bibitem[Wheeler et al.(2005)]{whe05} Wheeler, J. C., Akiyama, S., \& Williams, P. T. 2005, Ap\&SS, 298, 3
\bibitem[Wheeler et al.(2008)]{whe08} Wheeler, J. C., Maund, J.R., \& Couch, S.M. 2008, ApJ, 677, 1091
\bibitem[Willingale et al.(2002)]{will02} Willingale, R. Bleeker, J. A. M., van der Heyden, K. J., Kaastra, J. S., \& Vink, J. 2002, A\&A, 381, 1039
\bibitem[Willingale et al.(2003)]{will03} Willingale, R., Bleeker, J. A. M., van der Heyden, K. J., \& Kaastra, J. S. 2003, A\&A, 398, 1021
\bibitem[Winkler \& Kirschner (1985)]{wink85} Winkler, P. F., \& Kirshner, R. P. 1985, ApJ, 299, 981
\bibitem[Winkler et al. (1991)]{wink91} Winkler, P.F., Roberts, P.F., \& Kirshner R.P. 1991, in Supernovae: The Tenth Santa Cruz Summer Workshop in Astronomy and 
Astrophysics, ed. S.E. Woosley (New York:Springer), 652
\bibitem[Winkler \& Long (1997)]{wl97} Winkler, P. F., \& Long, K. S. 1997, ApJ, 491, 829
\bibitem[Woosley et al. (2002)]{woosley02} Woosley, S.E., Heger, A., \& Weaver, T.A. 2002, RvMP, 74, 1015
\bibitem[Young et al. (2006)]{young06} 	Young, P. A., Fryer, C. L., Hungerford, A., Arnett, D., Rockefeller, G., Timmes, F. X., Voit, B., Meakin, C., \& Eriksen, K. A. 2006, ApJ, 640, 891

\end{thebibliography}
\end{document}